\documentclass[11pt]{amsart}
\usepackage{fullpage}
\usepackage[foot]{amsaddr}
\usepackage{microtype}
\usepackage[OT1]{fontenc}
\usepackage{amsmath}
\usepackage{amssymb}
\usepackage{amsthm}
\usepackage{thmtools}
\makeatletter
\@ifundefined{newcounteralias}{}{\renewcommand\thmt@autorefsetup{\@xa\def\csname\thmt@envname autorefname\@xa\endcsname\@xa{\thmt@thmname}}}
\makeatother
\usepackage[linesnumbered,boxed,ruled,vlined]{algorithm2e}
\usepackage{algpseudocode}
\usepackage{enumitem}
\usepackage{xifthen}
\usepackage{xspace}

\usepackage[margin=1cm]{caption}
\usepackage{subfig}

\usepackage[thinlines]{easytable}

\usepackage[bookmarks=true,hypertexnames=false]{hyperref}
\hypersetup{hidelinks}

\usepackage{pgfplots}
\pgfplotsset{compat=1.16}
\usepackage{tikz}
\usetikzlibrary{arrows,arrows.meta,backgrounds,calc,fit,decorations.pathreplacing,decorations.markings,shapes.geometric}

\tikzstyle{internal} = [draw, fill, shape=circle]
\tikzstyle{external} = [shape=circle]
\tikzstyle{square}   = [draw, fill, rectangle]
\tikzstyle{triangle} = [draw, fill, regular polygon, regular polygon sides=3, inner sep=3pt]
\tikzstyle{pentagon} = [draw, fill, regular polygon, regular polygon sides=5, inner sep=2pt, minimum size=14pt]
\tikzset{every fit/.append style=text badly centered}

\usetikzlibrary{positioning,chains,fit,shapes,calc}
\usetikzlibrary{trees}
\usetikzlibrary{decorations.pathreplacing}
\usetikzlibrary{decorations.pathmorphing}
\usetikzlibrary{decorations.markings}
\tikzset{>=latex}

\usepackage{ifthen}

\usepackage{cleveref}

\usepackage[textsize=tiny]{todonotes}

\usepackage[normalem]{ulem}

\usepackage{mleftright}

\usepackage{cool}
\Style{DSymb={\mathrm d},DShorten=true,IntegrateDifferentialDSymb=\mathrm{d}}

\newcommand{\Ex}{\mathop{\mathbb{{}E}}\nolimits}
\renewcommand{\E}[2][]{\ifthenelse{\isempty{#1}}
  {\Ex\left[#2\right]}
  {\Ex_{#1}\left[#2\right]}}
\renewcommand{\Pr}{\mathop{\mathrm{Pr}}\nolimits}

\def\*#1{\mathbf{#1}}
\def\+#1{\mathcal{#1}}
\def\-#1{\mathrm{#1}}
\def\=#1{\mathbb{#1}}
\def\^#1{\mathbb{#1}}

\newcommand{\norm}[2]{\ensuremath{\Vert #2 \Vert_{#1}}}
\newcommand{\abs}[1]{\ensuremath{\left\vert#1\right\vert}}

\newcommand{\dist}{\operatorname{dist}}

\newcommand{\defeq}{:=}

\newtheorem{theorem}{Theorem}

\newtheorem{lemma}[theorem]{Lemma}

\newtheorem{proposition}[theorem]{Proposition}

\theoremstyle{definition}

\theoremstyle{remark}

\newtheorem*{remark*}{Remark}

\crefname{theorem}{Theorem}{Theorems}
\crefname{observation}{Observation}{Observations}
\crefname{claim}{Claim}{Claims}
\crefname{condition}{Condition}{Conditions}
\crefname{algorithm}{Algorithm}{Algorithms}
\crefname{property}{Property}{Properties}
\crefname{example}{Example}{Examples}
\crefname{fact}{Fact}{Facts}
\crefname{lemma}{Lemma}{Lemmas}
\crefname{corollary}{Corollary}{Corollaries}
\crefname{definition}{Definition}{Definitions}
\crefname{remark}{Remark}{Remarks}
\crefname{proposition}{Proposition}{Propositions}
\crefname{equation}{equation}{equations}
\crefname{enumi}{Case}{Case}
\creflabelformat{enumi}{(#2#1#3)}
\numberwithin{equation}{section}

\title[Deterministic Approximate Counting Beyond Bounded Degree]{Deterministic Approximate Counting\\
Beyond Bounded Degree}
\author{Yitong Yin}
\address[Yitong Yin]{School of Computer Science, Nanjing University, Nanjing, China}
\email{yinyt@nju.edu.cn}
\author{Xinyuan Zhang}
\address[Xinyuan Zhang]{School of Computer Science, Georgia Institute of Technology, Atlanta, GA, USA}
\email{normalvra@gmail.com}

\begin{document}



\begin{abstract}
We develop a method for deterministic approximate counting on graphs without degree restrictions.
For the hardcore and zero-field Ising models, our method improves the previous deterministic running-time bounds from quasipolynomial to polynomial time on general graphs.
In their tree-uniqueness regimes with any fixed slack $\delta\in(0,1)$, our algorithms achieve relative error $\epsilon$ in $(n/\epsilon)^{O_\delta(1)}$ time on any $n$-vertex graph.
The same bound also holds for proper $q$-colorings when $q\ge(2+\delta)\Delta$, where the maximum degree $\Delta$ may grow with $n$.
For the monomer--dimer model at fixed activity, our method improves the previous subexponential bound to quasipolynomial time.

Our method first gives a deterministic FPTAS under a general high-temperature condition by expanding the vertex-deletion recursion as geometric series and compressing the resulting sums of products.
We then rely on zero-freeness of the partition functions to extend these estimates to the target regimes by analytic extrapolation.
\end{abstract}

\maketitle

\section{Introduction}

A foundational question in the theory of computing is when randomness can speed up computation.
Approximate counting has long been a testing ground for this question, and many of its benchmark problems concern the partition functions of spin systems on graphs.
In their pioneering papers, Jerrum and Sinclair used Markov chain Monte Carlo (MCMC) to give fully polynomial-time randomized approximation schemes (FPRASes) for counting matchings in general graphs~\cite{JS89} and for the ferromagnetic Ising partition function~\cite{JS93}.
Rapidly mixing chains also yielded approximation schemes for the hardcore partition function~\cite{LV97,ALO20} and for counting proper graph colorings~\cite{Jer95,Vig00}.
The same problems have shaped the development of deterministic approximate counting.
In a seminal paper, Weitz used correlation decay to approximate the hardcore partition function up to the tree-uniqueness threshold~\cite{Wei06}. The same principle led to deterministic approximate counting results for matchings~\cite{BGKNT07}, antiferromagnetic two-spin systems including the Ising model~\cite{SST12,LLY13}, and list colorings~\cite{GK07,LY13}.
Another major approach combines Barvinok's zero-free interpolation method~\cite{Bar16} with Patel and Regts' algorithm for computing low-order coefficients from counts of small connected induced subgraphs~\cite{PR17}.
For graphs of fixed maximum degree $\Delta$, Liu, Sinclair and Srivastava used this approach to give a deterministic approximation scheme for counting proper $q$-colorings whenever $q\ge 2\Delta$~\cite{LSS19}.

A fundamental challenge in deterministic approximate counting is to control how the running time depends on the maximum degree $\Delta$ of the underlying graph.
For the benchmark problems above, randomized algorithms for approximate counting and sampling can run in polynomial time with exponents independent of $\Delta$~\cite{JS89,Jer95,CFYZ21,CFYZ22}.
For deterministic counting algorithms, a recurring obstacle is the cost of enumerating local configurations.
Correlation-decay algorithms explore paths in a computation tree~\cite{Wei06,BGKNT07}, and algorithms based on zero-free interpolation~\cite{PR17,LSS19} or cluster expansions~\cite{HPR20,BCHPT23} enumerate small connected subgraphs or clusters.
Even when structures of size $L=O(\log(n/\epsilon))$ suffice, their enumeration can cost $\Delta^{O(L)}=(n/\epsilon)^{O(\log\Delta)}$
per root.
The degree also affects the cost of enumerating the possible random choices of a truncated local sampler~\cite{FGWWY25} or constructing a linear program over a truncated coupling tree~\cite{CFGZZ25}.

Monte Carlo methods can estimate the collective contribution of many configurations from samples. 
The challenge is to obtain such an estimate deterministically when even the local configurations are too numerous to enumerate.
This leads to our central question:
\begin{quote}
\emph{Can deterministic approximate counting be efficient without degree restrictions?}
\end{quote}

We give an affirmative answer to this question.
As applications of our general method, we obtain deterministic fully polynomial-time approximation schemes (FPTASes) for the following benchmark spin systems, with running-time exponents independent of the maximum degree $\Delta$.
This improves deterministic quasipolynomial bounds for the hardcore and zero-field Ising models~\cite{Wei06,LSS19Ising}, and exponential bounds for colorings near $q=2\Delta$ when $\Delta=\Theta(n)$~\cite{LSS19,CFGZZ25,BBR26}.
\begin{theorem}\label{thm:main}
Fix $\delta\in(0,1)$. For each of the following models, there is a deterministic algorithm that approximates the partition function on any finite simple $n$-vertex graph $G$ of maximum degree $\Delta=\Delta(G)\ge3$ to any prescribed relative error $\epsilon\in(0,1)$ in
$(n/\epsilon)^{O_\delta(1)}$ time.
\begin{enumerate}
    \item The hardcore model with fugacity $0\le\lambda \le (1-\delta)\lambda_c(\Delta)$, where $\lambda_c(\Delta) = \frac{(\Delta-1)^{\Delta-1}}{(\Delta-2)^\Delta}$.
    \item The zero-field Ising model with monochromatic edge weight $\beta \in \bigl[\frac{\Delta-2+\delta}{\Delta-\delta},\frac{\Delta-\delta}{\Delta-2+\delta}\bigr]$ and bichromatic edge weight~1.
    \item Proper $q$-colorings for any integer $q\ge(2+\delta)\Delta$.
\end{enumerate}
\end{theorem}

Here $\lambda_c(\Delta)$ is the uniqueness threshold on the infinite $\Delta$-regular tree, and the stated Ising interval reaches the tree-uniqueness boundary with slack $\delta$.
For proper colorings, $q\ge(2+\delta)\Delta$ ensures Dobrushin's uniqueness condition with fixed slack.


Our second result gives a deterministic quasipolynomial-time approximation for the monomer--dimer partition function on general graphs at every fixed positive activity, improving the previous subexponential-time bound at fixed relative error~\cite{BGKNT07}.

\begin{theorem}\label{thm:match}
There is a deterministic algorithm that approximates the monomer--dimer partition function $Z_G(\lambda)=\sum_{M\text{ matching in }G}\lambda^{|M|}$ on any finite simple graph $G$ with $n\ge2$ vertices and edge activity $\lambda\ge0$ to any prescribed relative error $\epsilon\in(0,1)$ in $n^{O(\lceil\lambda\rceil\log^2(n/\epsilon))}$ time, where the implicit constant is absolute.
\end{theorem}


\begin{remark*}[Computation model]
The running times in \Cref{thm:main,thm:match} count arithmetic operations and comparisons over the reals.
For rational numerical inputs of at most $b$ bits each, including $\epsilon$, the corresponding bit complexities satisfy the same asymptotic bounds multiplied by $\operatorname{poly}(b)$.
\end{remark*}

\subsection{Technical Overview}
The proof of~\Cref{thm:main} proceeds in two steps.
First, we develop an FPTAS for spin systems satisfying the high-temperature assumptions of~\Cref{thm:high-temp-spin}, with a running-time exponent independent of $\Delta$.
Second, we rely on zero-freeness of the partition functions to extend these estimates to the parameter regimes of~\Cref{thm:main} by analytic extrapolation (\Cref{sec:extrapolation}).

The high-temperature algorithm expands the denominators in the vertex-deletion recursion as geometric series.
We represent the retained coefficients of the normalized insertion ratios as signed sums of products of bounded functions of consecutive history labels.
The high-temperature condition bounds the total absolute coefficient weight, allowing each sum to be compressed to a short list with controlled $L^2$ error under an auxiliary Markov chain on histories.
We select the terms deterministically using inner products between the product functions, which can be computed by dynamic programming over the states of the chain.
The chain is chosen so that the message updates do not amplify the coefficient errors under the history law.
The compression errors therefore add over the recursion depths, while compressing intermediate sums and products keeps the stored lists polynomial in size.

For the extrapolation step, keeping the running-time exponent
independent of $\Delta$ requires zero-free neighborhoods of uniform
radius after parameter rescaling.
For hardcore and Ising, we combine fixed-degree and large-degree
zero-free results.
For colorings, we start in the high-temperature regime a fixed distance
below edge weight~$1$, using the uniform neighborhood from
Appendix~\ref{app:potts-tube}.

For matchings, the messages are probabilities that vertices are unmatched, and the references approximate a fixed point of the monomer recursion.
At small positive activities, a weighted stopping rule controls the effects of deletions and the error from terminal messages.
We expand the relative errors around the reference messages and use the same compression framework to compute their low-degree coefficients in an auxiliary variable.
Evaluating the resulting polynomials gives the monomer probabilities and hence $\log Z_G$ at small activities.
The Heilmann--Lieb theorem~\cite{HL72} then allows extrapolation in the activity to the target value.
Tracking the coefficient precision and the activity extrapolation gives the quasipolynomial running time in~\Cref{thm:match} (\Cref{sec:matching}).

\subsection{Related Work}\label{sec:related-work}

Deterministic approximate counting for spin systems has developed through several overlapping approaches, including correlation decay~\cite{Wei06,BGKNT07,LLY13}, zero-free interpolation~\cite{Bar16,PR17,LSS19}, polymer and contour expansions~\cite{HPR20,BCHPT23}, derandomization of local samplers~\cite{FGWWY25}, and linear programming based on couplings~\cite{Moi19,CFGZZ25}.

For the hardcore and zero-field Ising regimes, \Cref{thm:main} improves
the $n^{O_\delta(\log n)}$ time bound obtained by truncating
Weitz's self-avoiding walk tree~\cite{Wei06,LSS19Ising} to polynomial
time at fixed relative error.
For proper colorings, Lu and Yin~\cite{LY13} obtained a quasipolynomial-time algorithm when $q\ge2.5808\Delta+1$.
For $\Delta=\Theta(n)$ and $q=\lceil(2+\delta)\Delta\rceil$ with small fixed $\delta>0$, the published bounds~\cite{LSS19,CFGZZ25,BBR26} give no improvement over exact counting~\cite{BHK09}.

For the monomer--dimer model, Bayati, Gamarnik, Katz, Nair, and
Tetali~\cite{BGKNT07} gave an FPTAS on bounded-degree graphs and a
deterministic algorithm taking $\exp(O(\sqrt{\lambda n}\log^2 n))$
time on general graphs at fixed positive activity $\lambda$ and
fixed relative error.
\Cref{thm:match} improves the subexponential worst-case bound to
quasipolynomial time.
Barvinok~\cite{Bar19} used the Heilmann--Lieb zero bound to approximate
the total number of matchings from exact counts of small matchings.
Our interpolation instead uses estimates of the log partition
function at small positive activities.

One way to accommodate unbounded maximum degree is to bound graph growth or require local sparsity.
Sinclair, Srivastava, \v{S}tefankovi\v{c} and Yin~\cite{SSSY17} give
FPTASes on graph families of bounded connective constant, which controls
self-avoiding walk growth, for the hardcore model below the associated
uniqueness threshold and for the monomer--dimer model at fixed activity.
Contraction methods give FPTASes for spin systems under growth or
sparsity restrictions~\cite{YZ16,LiYang26}, while
Efthymiou~\cite{Eft20} gives a deterministic polynomial-time
approximation to the coloring free energy on sparse random graphs
with sufficiently many colors.

On arbitrary graphs, Li, Lu and Yin~\cite{LLY13} give an FPTAS for
fixed antiferromagnetic two-spin parameters with $\beta<1<\gamma$
strictly inside the uniqueness region for every infinite regular tree.
Their \emph{computationally efficient correlation decay} exploits
increasingly strong contraction at sufficiently high degrees, yielding
a running-time exponent depending on the parameters but
independent of $\Delta$.

For the Sherrington--Kirkpatrick model, Harbuzova, Lepsveridze,
Majid and Moitra~\cite[Theorem~1.1]{HLMM26} and Schmidhuber and
Zlokapa~\cite[Theorem~31]{SZ26} give deterministic polynomial-time
approximations of the partition function at fixed inverse temperature
$0\le\beta<1$, with probability tending to one over Gaussian interactions.
The first combines color coding of a high-temperature expansion with
zero-freeness~\cite{BHLLR25}; the second uses the cavity method.
Yi~\cite[Theorem~1.2]{Yi26} gives a deterministic FPTAS for zero-field
Ising models with couplings of magnitude $O(1/n)$ under a one-sided
spectral condition with fixed slack, using Gaussian resummation and
truncation.
This covers our Ising regime when $\Delta=\Theta(n)$, whereas
\Cref{thm:main} is uniform in $\Delta$.

\subsection*{AI Disclosure}
GPT-6 Astra suggested the main idea of the spin-system argument.
GPT-6 Astra, GPT-6.2 Astra and Gemini-4 Argon assisted in the preparation of this manuscript.

\section{High-temperature FPTASes}\label{sec:high-temp}
Let $q\ge2$ and let $G=(V,E)$ be a finite simple graph on $n\ge1$ vertices with maximum degree at most $\Delta$, where $\Delta\ge3$.
Each vertex $v$ has an external field vector $\lambda_v\in\mathbb R_{\ge0}^q$ with $\norm{1}{\lambda_v}=1$.
Write $A_{vu}(c,b)$ for the edge weight when $v$ has spin $c$ and $u$ has spin $b$, so $A_{uv}(b,c)=A_{vu}(c,b)$.
Under the product measure $\pi$, vertex $v$ draws its spin from the distribution $\lambda_v$.
The partition function is 
\[
Z_G=\E[\sigma\sim\pi]{\prod_{uv\in E} A_{vu}(\sigma_v,\sigma_u)}.
\]
\begin{theorem}\label{thm:high-temp-spin}
    We impose the following restrictions on the spin system.
    \begin{itemize}
    \item (Edge weights) $0 \le A_e(c_1,c_2) \le 1$ for every edge $e$ and all $c_1,c_2 \in [q]$.
    \item (High-temperature) For every vertex $v$ and spin $c \in [q]$, it holds
    \begin{align}\label{eq:high-temp}
        \sum_{u\in N_G(v)} \E[\sigma \sim \pi]{1 - A_{vu}(c,\sigma_u)} \le \frac{1}{100}.
    \end{align}
    \end{itemize}
    For every $\epsilon\in(0,1)$, there is a deterministic algorithm approximating the partition function $Z_G$ to relative error $\epsilon$ using $\mathrm{poly}(n,q,\epsilon^{-1})$ arithmetic operations and comparisons.
\end{theorem}

\subsection{Outline}\label{subsec:outline}
We first outline the algorithm for the hardcore model and then describe its extension to the spin systems of~\Cref{thm:high-temp-spin}.
The starting point is the tree recursion~\cite{Wei06}. For a graph $G=(V,E)$, let $R_{G,v}$ be the marginal ratio at vertex $v$, i.e., the probability that $v$ is occupied divided by the probability that it is unoccupied. The tree recursion reads
\begin{align*}
    R_{G,v} = \lambda \prod_{i=1}^{d} \frac{1}{1+R_{G_i,v_i}},
\end{align*}
where $d$ is the degree of $v$ in $G$, $v_1,v_2,\ldots,v_d$ are its neighbors in a fixed total order on $V$, and $G_i$ is obtained from $G$ by removing $v,v_1,\ldots,v_{i-1}$.
Unrolling this recursion from $G$ and $v$ identifies each call with a history $h=(w_0,\ldots,w_\ell)$ where $w_0=v$ and $w_\ell=w$.
The marginal ratio $R_h$ of this particular call $h$ is thus
\begin{align}\label{eqn:R-recursion}
    R_h = \lambda \prod_{u \in N_G(w)} \frac{1}{1+I_h(u) R_{h \parallel u}} = \lambda \prod_{u \in N_G(w)} \sum_{t \ge 0} (-I_h(u) R_{h \parallel u})^t.
\end{align}
where $I_h(u)$ denotes the indicator that vertex $u$ is alive at history $h$, and the infinite sum absolutely converges since the marginal ratio $R \le \lambda \ll 1$.

Storing the full dynamic programming table is costly since the number of histories can grow as $\Delta^{\Theta(\ell)}$ at depth $\ell$. 
The solution here is to store approximations $\widehat{R_h}$ that exhibit succinct representations.
A key motivating example is the indicator $I_h(u)$, which already factors along the history as
\begin{align*}
    I_h(u) = \prod_{j=0}^{\ell-1} \chi_{w_j,w_{j+1}}(u),
\end{align*}
where $\chi_{a,b}(u)$ indicates that $u \neq a$ and either $u \notin N_G(a)$ or $u \ge b$.
Since each factor depends only on two consecutive vertices along $h$, this suggests representing $\widehat{R_h}$ as a short linear combination of local-factor products:
\begin{align}\label{eqn:Rh}
    \widehat{R_h} = \sum_{k=1}^M a_k \prod_{j=0}^{\ell-1} f_{k,j}(w_j,w_{j+1}).
\end{align}

The main challenge is to maintain the form~\eqref{eqn:Rh} throughout the recursion while keeping the error controlled.
Although the infinite series in~\eqref{eqn:R-recursion} can simply be truncated, a more serious computational issue is that adding or multiplying two expressions of the form~\eqref{eqn:Rh} can produce $2M$ or $M^2$ terms, respectively.
Such growth quickly becomes prohibitive across repeated operations.
Our solution is to apply a deterministic compression procedure, introduced in~\Cref{subsec:alg}, to reduce the number of terms back to $M$ while incurring only a controlled error.

Combining these ingredients gives the following algorithm for estimating $R_{G,v}$.
The dynamic programming table is indexed by the depth $\ell$ and the last vertex $w$ of a history.
Each entry stores a weighted sum of products in the form~\eqref{eqn:Rh}, approximating $R_h$ over histories with that depth and endpoint.
We initialize the entries at depth $n$ to the constant $\lambda$ and fill the table backwards.
For each neighbor $u$ of $w$, we fix the last transition $(w,u)$ in the child representation at $(\ell+1,u)$, obtaining an approximation to $R_{h\parallel u}$ as a function of the parent history $h$.
We multiply this approximation by $I_h(u)$ and substitute it into the truncated geometric series for $u$ in~\eqref{eqn:R-recursion}.
We then multiply the resulting factors and scale by $\lambda$, compressing after each multiplication.
The entry at depth zero and vertex $v$ gives the estimate of $R_{G,v}$.

For child entries with $|f_{k,j}|\le1$ and coefficient weight at most $2\lambda$, the update weight and the magnitude of each child derivative in~\eqref{eqn:R-recursion} are at most $\lambda(1-2\lambda)^{-\Delta-1}\le2\lambda$ when $\lambda\Delta$ is small.
With at most $\Delta$ children, Jensen's inequality along simple random walk gives
\[
    E_\ell\le2\lambda\Delta\,E_{\ell+1}+e_\ell,
\]
where $E_\ell$ is the $L^2$ error at depth $\ell$ under the walk's history law.
Since $E_n=0$ and $2\lambda\Delta<1$, iterating this inequality bounds the root error by the sum of local errors $e_\ell$.

For the general spin system, we introduce a vacancy spin $\star$ at each vertex. For $t\in[0,1]$, set $\lambda_v^\star(\star)=1-t$ and $\lambda_v^\star(c)=t\lambda_v(c)$ for $c\in[q]$, and extend each interaction matrix by $A_e^\star(\star,c)=A_e^\star(c,\star)=1$ for all $c\in[q]\cup\{\star\}$. Let $Z_G(t)$ be the resulting partition function, so that $Z_G(0)=1$ and $Z_G(1)$ is the original partition function.

For $0\le t<1$, consider an instance $H$ with color weights $t\lambda_v^H(c)$ and vacancy weight $1-t$. The all-vacant configuration has positive weight. For $v\in V(H)$ and $c\in[q]$, we define the ratio
\begin{align}\label{eq:def-R}
    R_{H,v}(c;t):=\frac{Z_H(t;\sigma_v=c)}{Z_{H-v}(t)},
\end{align}
where the restricted partition function includes the external field at $v$.
Order the neighbors $v_1,\ldots,v_d$ by the same total order on $V$. Inserting the interactions incident with $v$ one at a time gives the exact recursion
\begin{align}\label{eq:recursive-general}
    R_{H,v}(c;t)
    =t\lambda_v^H(c)\prod_{i=1}^d
    \left(1-
    \frac{\sum_{b=1}^q (1-A_{vv_i}(c,b))R_{H_i^c,v_i}(b;t)}
    {1-t+\sum_{b=1}^q R_{H_i^c,v_i}(b;t)}\right),
\end{align}
where $H_i^c$ is obtained from $H-v$ by replacing $\lambda_{v_j}^H(b)$ with $\lambda_{v_j}^H(b)A_{vv_j}(c,b)$ for every $j<i$ and $b\in[q]$. The $i$th factor is the expected weight of the next interaction under the spin distribution at $v_i$ in $H_i^c$.

For the formal expansion at $t=0$, write $z=t/(1-t)$ and $X_i(b)=R_{H_i^c,v_i}(b;t)/t$.
The same geometric expansion as in~\eqref{eqn:R-recursion} gives the $i$th neighbor factor in~\eqref{eq:recursive-general} as
\[
    1-\frac{z\sum_b(1-A_{vv_i}(c,b))X_i(b)}{1+z\sum_b X_i(b)}
    =1+\sum_{k\ge1}(-z)^k
       \left(\sum_b(1-A_{vv_i}(c,b))X_i(b)\right)
       \left(\sum_b X_i(b)\right)^{k-1}.
\]
As in the hardcore model, we use a dynamic programming table indexed by depth and the last vertex--spin pair to represent the normalized ratios as weighted sums of products with polynomial coefficients in $z$.
To update an entry, we fix the last transition in each child representation, sum over child spins, and multiply the truncated neighbor factors, compressing coefficient by coefficient after each multiplication.

For an induced subgraph $H$ with its original external fields, summing over the spin at $v$ gives
\begin{align}\label{eq:deletion-ratio}
    Q_{H,v}(t):=\frac{Z_H(t)}{Z_{H-v}(t)}=1-t+\sum_{c=1}^q R_{H,v}(c;t).
\end{align}
We estimate the low-degree coefficients of these ratios and evaluate the resulting polynomials.

\subsection{Algorithm and analysis}\label{subsec:alg}
We first give the compression routine used in the dynamic program.

\noindent\begin{algorithm}[H]
    \SetVlineSkip{1pt}
    \caption{Compression of a weighted sum}
    \label{alg:compress-suffix}
    \SetKwInOut{Input}{Input}
    \SetKwInOut{Output}{Output}
    \SetKwInOut{Parameter}{Parameters}
    \textsc{Compress} $((a_i,g_i)_{i=1}^p,\mu,M)$\;
    \Input{Real coefficients $a_i$ with $\sum_{i=1}^p |a_i|\le1$ and functions $g_i:\Omega\to[-1,1]$}
    \Parameter{An oracle for $\langle g_i,g_j\rangle_{L^2(\mu)}$ under a probability measure $\mu$, and an integer $M\ge1$}
    \Output{An average $\widehat G$ of $M$ signed input functions or zeros}
    $w_i\gets |a_i|$, \quad $\widetilde g_i\gets\operatorname{sgn}(a_i)g_i$ for $1\le i\le p$\;
    $w_0\gets1-\sum_{i=1}^p w_i$, \quad $\widetilde g_0\gets0$\;
    $G\gets\sum_{i=1}^p a_i g_i$, \quad $U\gets0$\;
    \For{$\ell=1,\ldots,M$}{
        Choose an index $i$ with $w_i>0$ minimizing $\|U+\widetilde g_i-\ell G\|_{L^2(\mu)}$\;
        $U\gets U+\widetilde g_i$\;
    }
    \Return $\widehat{G} = U/M$\;
\end{algorithm}

\begin{proposition}\label{lem:compress}
For $G=\sum_i a_i g_i$, the algorithm returns an average $\widehat G$
of $M$ functions from $\{\widetilde g_0,\ldots,\widetilde g_p\}$ satisfying
\[
    \|\widehat G-G\|_{L^2(\mu)}\le M^{-1/2}.
\]
For arbitrary coefficients with $B=\sum_i|a_i|>0$, the output of
\Cref{alg:compress-suffix} on coefficients $a_i/B$, rescaled by $B$,
has $L^2(\mu)$ error at most $B/\sqrt M$.
For $B=0$, the zero function is exact.
\end{proposition}
\begin{proof}
    Draw an index $I$ with $\Pr(I=i)=w_i$ for $0\le i\le p$, so $\mathbb E_I[\widetilde g_I]=G$. Write $\|\cdot\|_2$ for the $L^2(\mu)$ norm. At iteration $\ell$, put $R=U-(\ell-1)G$. Since $\mathbb E_I[\widetilde g_I-G]=0$ and $\|\widetilde g_I\|_2\le1$,
    \[
        \mathbb E_I\|R+\widetilde g_I-G\|_2^2
        =\|R\|_2^2+\mathbb E_I\|\widetilde g_I-G\|_2^2
        \le\|R\|_2^2+1.
    \]
    The minimizing index attains at most this average. Starting from $U=0$, we obtain $\|U-MG\|_2^2\le M$ after $M$ iterations. 
    Dividing by $M^2$ proves the error bound. The general case follows by scaling.
\end{proof}




We first estimate the low-degree coefficients of the deletion ratio in~\eqref{eq:deletion-ratio}.

\begin{lemma}\label{lem:spin-seed}
Under the assumptions of~\Cref{thm:high-temp-spin}, given $v\in V(G)$, $D\ge1$, and $\eta\in(0,1/2)$, a deterministic algorithm returns a polynomial $\widehat Q_{G,v}(t)$ of degree at most $D$ with $\sum_{j=0}^D16^{-j}|[t^j](\widehat Q_{G,v}-Q_{G,v})|\le\eta$, using $\operatorname{poly}(n,q,D,\eta^{-1})$ arithmetic operations and comparisons.
\end{lemma}

\begin{proof}
To estimate $Q_{G,v}(t)=1-t+\sum_c R_{G,v}(c;t)$, we compute the low-degree coefficients of $R_{G,v}(c;t)$ in $z = t/(1-t)$ using dynamic programming.

Fix a color $c_0\in[q]$ and set the root label to $s_0=(v,c_0)$; the dynamic programming table has $n+1$ rows and $nq$ columns.
The rows are indexed by history depth $\ell=0,\ldots,n$, and the columns by the final vertex--spin label $s\in V\times[q]$.
Let $Y_{\ell,s}$ be the exact message as a function of histories $h=(s_0,\ldots,s_\ell)$ with $s_\ell=s$. Using the representation in~\Cref{subsec:outline}, entry $(\ell,s)$ stores an approximation $\widehat Y_{\ell,s}$ of the form
\begin{equation}\label{eq:history-list}
    \widehat Y_{\ell,s}(h)=\sum_i {a_i(z)}\prod_{j=1}^{\ell}f_{i,j}(s_{j-1},s_j),
    \qquad |f_{i,j}|\le1.
\end{equation}
Each $a_i(z)$ has degree at most $D$ with $\sum_i \|a_i\|_\zeta \le 2$, where $\|P\|_\zeta=\sum_j \zeta^j|[z^j]P|$ and $\zeta=1/15$.

For an entry $(\ell,s)$ with $s=(v,c)$, write $h\parallel(u,b)$ for the history obtained by appending $(u,b)$ to $h$. The idealized recurrence from~\Cref{subsec:outline} reads, modulo $z^{D+1}$, as
\begin{equation}\label{eq:spin-message-update}
\begin{aligned}
    Y_{\ell,s}(h)&=L_{\ell,s}(h)\prod_{u\in N_G(v)}F_u(h),\\
    F_u(h)&=1+I_u(h)\sum_{k\ge1}(-z)^k
          \left(\sum_b(1-A_{vu}(c,b))\lambda_u(b)
                 Y_{\ell+1,(u,b)}(h\parallel(u,b))\right)\\
       &\hphantom{=1}\quad{}\times
          \left(\sum_b\lambda_u(b)Y_{\ell+1,(u,b)}(h\parallel(u,b))\right)^{k-1}.
\end{aligned}
\end{equation}
Here $L_{\ell,s}(h)$ is the inherited field multiplier at $v$, and $I_u(h)$ indicates whether $u$ remains present. For $s_j=(w_j,c_j)$, we have
\[
    L_{\ell,s}(h)=\prod_{\substack{0\le j<\ell:\ v\in N_G(w_j)\\v<w_{j+1}}}A_{w_jv}(c_j,c),
    \qquad I_u(h)=\prod_{j=0}^{\ell-1}\mathbf1_{\{w_j\ne u\}}.
\]
We initialize $Y_{n,s}(h)=\widehat Y_{n,s}(h)=L_{n,s}(h)$ as a single product and fill the remaining rows in decreasing order of depth.
The recurrence and terminal condition define the exact messages on all walks, including those with repeated vertices and zero-field spins. 
On a valid deletion history $h$ ending at $s=(w,c)$, the insertion ratio is given by $t\lambda_w(c)Y_{\ell,s}(h)$.

We now use \Cref{alg:compress-suffix} to implement this dynamic program in polynomial time.
Choose $\xi={\Theta(\eta/(n+1))}$ with a sufficiently small absolute constant and set $M=\lceil\xi^{-2}\rceil$.
When updating the cell $(\ell,s)$, write $\mathcal C$ for compression applied to the real coefficients at each degree separately, with $M$ terms per degree and the parent measure $\mu_{\ell,s}$ defined below, using the normalization in~\Cref{lem:compress}.
All polynomial operations in the dynamic program are performed modulo
$z^{D+1}$, discarding terms of degree greater than $D$ after every operation.

The terms with $k>D$ in the neighbor series vanish modulo $z^{D+1}$. We compute the powers of the second sum using
\[
    T_{u,0}(h)=1,\qquad
    T_{u,j}(h)=\mathcal C\!\left[
        h'\mapsto T_{u,j-1}(h')\sum_b\lambda_u(b)
        \widehat Y_{\ell+1,(u,b)}(h'\parallel(u,b))
    \right](h)
    \quad{(1\le j<D)}.
\]
The truncated neighbor factor $\widehat F_u(h)$ is then given by
\[
    \widehat F_u(h)=1+I_u(h)\sum_{k=1}^{{D}}(-z)^k
        \left(\sum_b(1-A_{vu}(c,b))\lambda_u(b)
               \widehat Y_{\ell+1,(u,b)}(h\parallel(u,b))\right)
        T_{u,k-1}(h).
\]
The approximation $\widehat Y_{\ell,s}(h)$ is computed similarly:
In neighbor order $u_1,\ldots,u_k$, set
\[
    \widehat U_0(h)=1,\qquad
    \widehat U_i(h)=\mathcal C\!\left[h'\mapsto\widehat U_{i-1}(h')\widehat F_{u_i}(h')\right](h), \qquad \widehat{Y}_{\ell,s}(h) = L_{\ell,s}(h) \widehat{U}_k(h).
\]
The normalized root estimate $\widehat Y_{0,s_0}$ is read as a polynomial in $z$ by evaluating it on the singleton history $(s_0)$.

The only missing piece of the algorithm is the compression measure $\mu_{\ell,s}$. Let $(S_j)_{j\ge0}$ be the Markov chain on vertex--spin labels with $S_0=s_0$ and transition matrix $K$ defined below. Write $\mu_\ell$ for the law of $(S_0,\ldots,S_\ell)$. For each cell $(\ell,s)$ with $\Pr(S_\ell=s)>0$, set
\[
    \mu_{\ell,s}=\operatorname{Law}\bigl((S_0,\ldots,S_\ell)\mid S_\ell=s\bigr).
\]

For each vertex--spin label $(v,c)$ and neighbor $u$ of $v$, set $d_u=\sum_b\lambda_u(b)(1-A_{vu}(c,b))$.
Define the neighbor weights $W_{(v,c),(u,b)}$ by
\begin{equation}\label{eq:spin-walk-weight}
    W_{(v,c),(u,b)}
    =2{\zeta}\lambda_u(b)\left(
        \frac{1-A_{vu}(c,b)}{1-2{\zeta}}
        +\frac{2{\zeta}d_u}{(1-2{\zeta})^2}\right)
    \qquad (u\in N_G(v)).
\end{equation}
By~\eqref{eq:high-temp}, each row sum of $W$ is at most $1/2$. We define the transition matrix $K$ by setting $K_{s,s'}=2W_{s,s'}$ for $s'\ne s$ and placing the remaining probability $1-2\sum_{s'\ne s}W_{s,s'}$ on the self-loop $K_{s,s}$.
All other entries of $W$ are zero.

We compute only cells of positive probability and omit zero-weight summands. 
When $d_u=0$, the neighbor factor is identically one, so we set $\widehat F_u=1$ directly.
When $d_u>0$, every child with $\lambda_u(b)>0$ has positive transition probability, so all required child cells are available.

We show by backward induction that each stored list in~\eqref{eq:history-list} satisfies $\sum_i\|a_i\|_\zeta\le2$. At depth $n$ each list is a single product of weight $1$. Truncated convolution is submultiplicative for both $\|\cdot\|_\zeta$ and the list weight. If the child lists have weight at most $2$, the two color sums have weights at most $2d_u$ and $2$, respectively. Compression does not increase weight, so the input and output of the $j$th power compression both have weight at most $2^j$. Since $\|z\|_\zeta=\zeta$, the list for $\widehat F_u$ has weight at most
\[
    1+2d_u\sum_{k=1}^D\zeta^k2^{k-1}\le1+\beta_u,
    \qquad \beta_u=\frac{2\zeta d_u}{1-2\zeta}.
\]
Each partial-product list has weight at most the product of the corresponding bounds $1+\beta_u$, since compression does not increase list weight. Multiplication by $L_{\ell,s}(h)$ only multiplies the local factors in each term by factors of absolute value at most $1$. Thus, by~\eqref{eq:high-temp}, the updated list satisfies
\[
    \sum_i\|a_i(z)\|_\zeta\le\prod_u(1+\beta_u)\le \exp\!\left(\sum_u\beta_u\right)<2.
\]
This completes the induction. Since $|f_{i,j}|\le1$, the triangle inequality in~\eqref{eq:history-list} gives
\[
    \|\widehat Y_{\ell,s}(h)\|_\zeta\le\sum_i\|a_i(z)\|_\zeta\le2.
\]
The same backward induction, using pointwise $\|\cdot\|_\zeta$ norms for the color sums and neighbor products, gives $\|[Y_{\ell,s}(h)]_{\le D}\|_\zeta\le2$ from the exact initialization.

It remains to analyze the time complexity and error of this algorithm.
Each table has $O(n^2q)$ cells, and each cell processes at most $n$ neighbors with $D$ series terms per neighbor and at most $M(D+1)$ terms per compressed list.
The pointwise product of two terms again factors along the history, so their conditional inner product is computed by a forward pass along the Markov chain, normalized by the endpoint probability.
These inner products determine the squared norms in~\Cref{alg:compress-suffix}, so the total cost is $\operatorname{poly}(n,q,D,\eta^{-1})$ arithmetic operations and comparisons.

For cells of zero probability, set $E_{\ell,s}=0$; otherwise let
\[
    E_{\ell,s} \defeq \left(\mathbb E_{h\sim\mu_{\ell,s}}
        \left[\|e_{\ell,s}(h;z)\|_\zeta^2\right]\right)^{1/2},
\]
where $e_{\ell,s}(h;z)=\widehat Y_{\ell,s}(h)-[Y_{\ell,s}(h)]_{\le D}$ is the difference between the stored polynomial and the degree-$D$ truncation of the exact message.
We claim that the row error $E_\ell\defeq(\sum_s\Pr(S_\ell=s)E_{\ell,s}^2)^{1/2}$ satisfies
\[
    E_n=0,\qquad E_\ell\le O(n\xi)+\tfrac12 E_{\ell+1}\quad(0\le\ell<n).
\]

The initialization gives $E_n=0$. At a cell $(\ell,s)$ with $\ell<n$, let $\widetilde Y_{\ell,s}$ be the exact update~\eqref{eq:spin-message-update} modulo $z^{D+1}$ on the stored children. Writing $h=(s_0,\ldots,s_\ell)$, the triangle inequality gives
\[
    E_\ell\le
    \left(\mathbb E_{h\sim\mu_\ell}
        \bigl[\|\widehat Y_{\ell,s_\ell}(h)-\widetilde Y_{\ell,s_\ell}(h)\|_\zeta^2\bigr]\right)^{1/2}
    +\left(\mathbb E_{h\sim\mu_\ell}
        \bigl[\|\widetilde Y_{\ell,s_\ell}(h)-[Y_{\ell,s_\ell}(h)]_{\le D}\|_\zeta^2\bigr]\right)^{1/2},
\]
where we call the first term the local error and the second the error propagated from the children.

We first bound the local error term by applying the scaled bound in~\Cref{lem:compress} at each degree. For one compression at $(\ell,s)$, let $P$ be the input polynomial, $\widehat P=\mathcal C P$ the compressed polynomial, and $B_j$ the weight of the degree-$j$ input list. The coefficient error satisfies
\[
    \bigl\|[z^j](\widehat P-P)\bigr\|_{L^2(\mu_{\ell,s})}
    \le \frac{B_j}{\sqrt M}.
\]
By the definition of $\|\cdot\|_\zeta$ and Minkowski's inequality,
\[
    \left(\mathbb E_{h\sim\mu_{\ell,s}}
      \bigl[\|\widehat P(h;z)-P(h;z)\|_\zeta^2\bigr]\right)^{1/2}
    \le \sum_{j=0}^D\zeta^j
      \bigl\|[z^j](\widehat P-P)\bigr\|_{L^2(\mu_{\ell,s})}
    \le \frac{B}{\sqrt M}\le B\xi,
\]
where $B=\sum_{j=0}^D\zeta^j B_j$ is the total input weight and $M\ge\xi^{-2}$.

Let $\delta_{u,j}$ be the error of $T_{u,j}$ relative to the exact $j$th power on the stored children, measured as above. Multiplication by the second sum at most doubles the error and the next compression adds at most $2^j\xi$, giving
\[
    \delta_{u,0}=0,\qquad
    \delta_{u,j}\le 2\delta_{u,j-1}+2^j\xi,
\]
which implies $\delta_{u,j}\le j2^j\xi$.
The error in each neighbor factor is at most
\[
    2\zeta d_u\sum_{j=0}^{D-1}\zeta^j\delta_{u,j}
    \le 2\zeta d_u\xi\sum_{j\ge0}j(2\zeta)^j
    =O(\zeta d_u\xi).
\]
Each partial-product compression adds at most $2\xi$. The remaining products in the telescoping sum have norm at most $2$, and multiplication by $L_{\ell,s}(h)$ does not increase the error. Thus the local error in a cell is at most
\[
    2\left(2|N_G(v)|\xi+O\!\left(\zeta\xi\sum_u d_u\right)\right)=O(n\xi).
\]

In the rational form of a neighbor factor, the inverse has norm at most $(1-2\zeta)^{-1}$ and the numerator before multiplication by $z$ has norm at most $2d_u$. The identity $A^{-1}-B^{-1}=A^{-1}(B-A)B^{-1}$ and the product bound $2$ bound the propagated error by the sum of the child errors weighted by~\eqref{eq:spin-walk-weight}. Since $W\le\tfrac12 K$ entrywise, for a history $h$ in the support of $\mu_{\ell,s}$ this gives
\[
\begin{aligned}
    \bigl\|\widetilde Y_{\ell,s}(h)
      -[Y_{\ell,s}(h)]_{\le D}\bigr\|_\zeta
    &\le \sum_{s'}W_{s,s'}
       \bigl\|e_{\ell+1,s'}(h\parallel s';z)\bigr\|_\zeta\\
    &\le \tfrac12\,\mathbb E\!\left[
      \bigl\|e_{\ell+1,S_{\ell+1}}
        (h\parallel S_{\ell+1};z)\bigr\|_\zeta
      \,\middle|\,(S_0,\ldots,S_\ell)=h\right].
\end{aligned}
\]
Jensen's inequality and averaging over $\mu_\ell$ bound the propagated error by $\tfrac12 E_{\ell+1}$. The uniform local bound contributes $O(n\xi)$ in the row norm, proving the claimed recurrence.

The recurrence gives singleton root error $O(n\xi)$. On valid histories the exact update agrees with~\eqref{eq:recursive-general}, with all neighbor indicators vanishing at a leaf.

Finally, substitute $z=t/(1-t)$ into each root polynomial and form
\[
    \widehat Q_{G,v}(t)
    =\left[
        1-t+t\sum_c\lambda_v(c)
        \widehat Y_{0,(v,c)}\!\left(\frac{t}{1-t}\right)
      \right]_{\le D}.
\]
The substitution preserves the truncation because $z(t)$ has zero constant term.
Since $\sum_{j=1}^D16^{-j}\le\zeta$, this substitution maps a
$\|\cdot\|_\zeta$ error to at most the same error in the weighted $t$-norm
of the lemma. Thus the returned polynomial has error $O(n\xi)\le\eta$.
\end{proof}

We bound the tails of the deletion-ratio series by controlling the effect of decreasing the external fields.
For a power series $f(t)$, write $\|f\|_*=\sum_{j\ge0}2^j|[t^j]f|$.

\begin{lemma}\label{lem:spin-field-stability}
Under the assumptions of~\Cref{thm:high-temp-spin}, let $H$ be an induced subgraph of $G$ with its original fields and let $0\le\mu_v(c)\le\lambda_v(c)$ for $v\in V(H)$ and $c\in[q]$.
Write $Z_H^\mu(t)$ for the partition function with color weights $t\mu_v(c)$ and vacancy weight $1-t$. Then
\begin{equation}\label{eq:spin-field-stability}
 \left\|\frac{Z_H^\mu(t)}{Z_H(t)}-1\right\|_*\le e^{5d}-1,\qquad
 d=\sum_{v\in V(H),\,c\in[q]}(\lambda_v(c)-\mu_v(c)).
\end{equation}
Moreover, for every $v\in V(H)$, the deletion ratio in~\eqref{eq:deletion-ratio} satisfies
\[
 \|Q_{H,v}-1\|_*\le2(e^{5/100}-1)<1/2 .
\]
\end{lemma}
\begin{proof}
Put $\alpha=1/100$. All quotients are initially formal series at zero, where every partition function equals one. The norm is submultiplicative.
We prove both bounds by induction on the size of $H$. The base case where $H$ is empty is immediate.

Write $\mu^{v,c}$ for the fields on $H-v$ obtained by multiplying each neighbor's field $\mu_u(b)$ by $A_{vu}(c,b)$. By~\eqref{eq:high-temp}, this reduces the total field mass on $H-v$ by at most $\alpha$. For the original fields, fixing $v$ to color $c$ gives $R_{H,v}(c;t)=t\lambda_v(c)F_c(t)$, where $F_c=Z_{H-v}^{\lambda^{v,c}}/Z_{H-v}$. The smaller-instance bound gives $\|F_c-1\|_*\le e^{5\alpha}-1$. By~\eqref{eq:deletion-ratio},
\[
 Q_{H,v}(t)-1=t\sum_c\lambda_v(c)(F_c(t)-1),\qquad
 \|Q_{H,v}-1\|_*\le2(e^{5\alpha}-1)<1/2 .
\]
Hence $\|Z_{H-v}/Z_H\|_*=\|1/Q_{H,v}\|_*\le2$ by the geometric series.

Now consider fields $\mu$ and $\nu$ which differ only at one vertex $v$, with $0\le\mu_v(c)\le\nu_v(c)=\lambda_v(c)$ and $\nu_u(c)\le\lambda_u(c)$ elsewhere. Summing over the spin at $v$ as in~\eqref{eq:deletion-ratio} and canceling the common vacancy term gives
\[
 \frac{Z_H^\nu-Z_H^\mu}{Z_H}
 =t\,\frac{Z_{H-v}}{Z_H}
    \sum_c(\nu_v(c)-\mu_v(c))\frac{Z_{H-v}^{\mu^{v,c}}}{Z_{H-v}} .
\]
Let $d$ and $d'$ be the deficits of $\mu$ and $\nu$, respectively. The deficit on $H-v$ after inserting the pins is at most $d'+\alpha$. Together with the induction hypothesis and $\|t\|_*=2$, this yields
\[
 \left\|\frac{Z_H^\nu-Z_H^\mu}{Z_H}\right\|_*
 \le4(d-d')e^{5(d'+1/100)}
 \le e^{5d}-e^{5d'}.
\]
The last inequality uses $4e^{5/100}\le5$ and $1+5(d-d')\le e^{5(d-d')}$. Restoring one vertex at a time and summing telescopes the right-hand side to $e^{5d}-1$, proving~\eqref{eq:spin-field-stability}.
\end{proof}

The norm bounds give absolute convergence of $Q_{H,v}$ and $1/Q_{H,v}$
on $|t|<2$; induction along a vertex-deletion order then shows that
$Z_H(t)\ne0$ there and identifies these series with the corresponding
partition-function ratios.

\begin{proof}[Proof of~\Cref{thm:high-temp-spin}]
Take the least $D\ge1$ with $2^{-D}\le\epsilon/(4n)$ and set $\eta=(\epsilon/(16n))16^{-D}$. Fix a vertex deletion order and write $Q$ for the deletion ratio of the current graph at each step. By~\Cref{lem:spin-field-stability}, the series for $Q$ converges absolutely at $t=1$, with $Q(1)\ge1/2$ and $\sum_{j>D}|[t^j]Q|\le2^{-D-2}$. Thus the product of the ratios at $t=1$ is $Z_G(1)$.

At each deletion step, apply~\Cref{lem:spin-seed} to obtain $\widehat Q$ and evaluate $\widehat Q(1)$ by summing its coefficients; the error satisfies
\[
    |\widehat Q(1)-Q(1)|\le16^D\eta+2^{-D-2}\le\frac{\epsilon}{8n}.
\]
Each ratio therefore has relative error at most $\epsilon/(4n)$, giving relative error at most $\epsilon$ for the product. Since $D=O(\log(n/\epsilon))$ and $\eta^{-1}$ is polynomial in $n/\epsilon$, the total cost is $\operatorname{poly}(n,q,\epsilon^{-1})$ arithmetic operations and comparisons.
\end{proof}

\begin{remark*}[Bit implementation]
Assume rational inputs of at most $b$ bits each.
In~\Cref{lem:spin-seed}, after each compression, round the local factors
and weighted coefficients $\zeta^j[z^j]a_i$ toward zero on a common
dyadic grid of mesh at most $\xi/[4(n2^D+(D+1)M)]$.
This preserves the list-weight bounds and adds at most $\xi/4$ per
compression, since the weight is at most $2^D$, each term has at most
$n$ factors, and there are at most $(D+1)M$ monomials.
The added error is absorbed by the choice of $\xi$.
All stored numbers have bit length polynomial in
$n,q,D,b,\log(1/\eta)$.
Taking common denominators along the forward computation of at most
$n$ steps gives the same bound for conditional inner products,
so squared norms can be compared exactly.
Hence~\Cref{lem:spin-seed} uses
$\operatorname{poly}(n,q,D,\eta^{-1},b)$ bit operations, and
\Cref{thm:high-temp-spin} uses
$\operatorname{poly}(n,q,\epsilon^{-1},b)$.
\end{remark*}

\section{Analytic Extrapolation}\label{sec:extrapolation}

We formalize the analytic extrapolation step that extends high-temperature estimates across a zero-free region.
For a real interval $I$ and $\rho>0$, write $B(I,\rho)=\{z\in\mathbb C:\dist(z,I)<\rho\}$.

\begin{lemma}\label{lem:extrapolation}
    Let $M,\delta>0$ and $r\in(0,1)$ be constants, and let $f$ be holomorphic on $B([0,1],\delta)$ with $|f|\le M$ there.
    Suppose an oracle takes $t\in[0,r]$ and $\eta>0$ and returns an estimate of $f(t)$ within additive error $\eta$.
    For every $\epsilon\in(0,1/2)$, a deterministic algorithm estimates $f(1)$ within additive error $\epsilon$ using
    $O_{M,\delta,r}(\log(1/\epsilon))$ oracle calls, each with tolerance $\eta=\epsilon^{O_{M,\delta,r}(1)}$, and
    $O_{M,\delta,r}(\log^2(1/\epsilon))$ further arithmetic operations.
\end{lemma}

\begin{proof}
    Choose an integer $N\ge\max\{2,1/r,2\mathrm e/\delta\}$ and partition $[0,1]$ into intervals
    $I_i=[(i-1)/N,i/N]$ for $1\le i\le N$.
    We first interpolate on $I_1\subseteq[0,r]$ and then successively extrapolate from $I_{i-1}$ to $I_i$, decreasing the degree at each stage.

    Let $d_1>\cdots>d_N\ge1$ be degrees to be chosen below.
    Put $J_1=I_1$ and $J_i=I_{i-1}$ for $i\ge2$.
    At stage $i$, use $d_i+1$ equally spaced nodes $x_{i,0},\ldots,x_{i,d_i}$ spanning $J_i$.
    The polynomial $P_1$ interpolates the oracle estimates at its nodes.
    For $i\ge2$, the polynomial $P_i$ interpolates the values $P_{i-1}(x_{i,j})$.
    Let $Q_i$ interpolate the exact values $f(x_{i,j})$ on the same grid.

    Set $\epsilon_i=\sup_{t\in I_i}|P_i(t)-f(t)|$ and $\epsilon_0=\eta$.
    The Hermite--Genocchi interpolation remainder and Cauchy's estimate give
    \[
        \sup_{t\in I_i}|f(t)-Q_i(t)|
        \le M\left(\frac{2}{N\delta}\right)^{d_i+1}
        \le M\mathrm e^{-(d_i+1)}.
    \]
    Each distance $|t-x_{i,j}|$ is at most $2/N$ for $t\in I_i$.
    Since the grid spacing is $1/(Nd_i)$, the sum of the absolute Lagrange weights is at most
    \[
        (2d_i)^{d_i}\sum_{j=0}^{d_i}\frac{1}{j!(d_i-j)!}
        \le(4\mathrm e)^{d_i}.
    \]
    Consequently,
    \[
        \epsilon_i\le M\mathrm e^{-(d_i+1)}
                     +\epsilon_{i-1}(4\mathrm e)^{d_i+1}.
    \]

    Put $L=\max\{2,\lceil\log_2(2M/\epsilon)\rceil\}$ and $d_i=4^{N-i}L-1$.
    Set $\eta=M\,32^{-(d_1+1)}$.
    Since $d_{i-1}+1=4(d_i+1)$ and $2(4\mathrm e^{-2})^k\le1$ for $k\ge2$, induction gives
    \[
        \epsilon_i\le2M\exp(-4^{N-i}L),\qquad 1\le i\le N.
    \]
    In particular, $P_N(1)$ has error at most $\epsilon$.
    Only the first stage calls the oracle, at $d_1+1=O_{M,\delta,r}(\log(1/\epsilon))$ nodes.
    With the Lagrange denominators precomputed, each evaluation costs linear time in the degree, giving $O_{M,\delta,r}(\log^2(1/\epsilon))$ arithmetic operations in total.
\end{proof}

For our applications, we use the following zero-free neighborhoods with uniform dependence on the degree.

\begin{lemma}[{\cite{PR19,BBP25}}]\label{lem:hardcore-tube}
    For every $\delta\in(0,1)$ there exists $\rho=\rho(\delta)>0$ such that, for every $\Delta\ge3$ and every finite simple graph $G$ of maximum degree at most $\Delta$, its independence polynomial satisfies
    \[
        Z_G^{\mathrm{hc}}(z)=\sum_{I\subseteq V(G):\,I\text{ independent}}z^{|I|}\ne0
        \qquad\text{for all }z\in B\bigl([0,(1-\delta)\lambda_c(\Delta)],\rho/\Delta\bigr).
    \]
\end{lemma}

\begin{lemma}[{\cite{LSS19Ising,PRS24}}]\label{lem:ising-tube}
    For every $\delta\in(0,1)$ there exists $\rho=\rho(\delta)>0$ such that the following holds for every $\Delta\ge3$ and every finite simple graph $G=(V,E)$ of maximum degree at most $\Delta$.
    Set $I_{\Delta,\delta}=\bigl[\frac{\Delta-2+\delta}{\Delta-\delta},\frac{\Delta-\delta}{\Delta-2+\delta}\bigr]$.
    With $m(\sigma)$ denoting the number of monochromatic edges, the zero-field Ising partition function satisfies
    \[
        Z_G^{\mathrm{Ising}}(z)=\sum_{\sigma\in\{+,-\}^V}z^{m(\sigma)}\ne0
        \qquad\text{for all }z\in B(I_{\Delta,\delta},\rho/\Delta).
    \]
\end{lemma}

\begin{proof}[Proof of~\Cref{lem:hardcore-tube,lem:ising-tube}]
Write $d=\Delta-1$.
For the hardcore model, let $\mathcal U_d$ be the common open zero-free region for graphs of maximum degree at most $d+1$.
By~\cite[Theorems~1.1 and~1.8]{BBP25}, every compact subset of a limiting open region $\mathcal U_\infty\supset[0,\mathrm e)$ is contained in $d\mathcal U_d$ for all sufficiently large $d$.
Choose a compact neighborhood of $[0,(1-\delta/2)\mathrm e]$ inside $\mathcal U_\infty$.
Since $d\lambda_c(\Delta)\to\mathrm e$, a fixed-width neighborhood of $[0,(1-\delta)d\lambda_c(\Delta)]$ lies in this compact set for all large $d$.
Scaling back gives the required tube of width $\rho/\Delta$.

For Ising, put
\[
    \phi_d(z)=\frac{d(z-1)}{z+1},
    \qquad r_d=\left(1-\frac{1}{\sqrt{2d}}\right)^2.
\]
By~\cite[Theorem~1.1]{PRS24}, the partition function is nonzero when $|\phi_d(z)|\le r_d$.
The map $\phi_d$ sends $I_{\Delta,\delta}$ onto $[-1+\delta,1-\delta]$, and $r_d\ge1-\delta/2$ for all sufficiently large $d$.
Moreover, $I_{\Delta,\delta}\subset[1/3,3]$, so $|\phi_d'|=O(d)$ on a fixed neighborhood of this interval.
Thus a sufficiently small $\rho=\rho(\delta)>0$ ensures that the $\rho/\Delta$ tube maps into $|\phi_d|\le1-\delta/2$.

For the remaining degrees, the fixed-degree results~\cite[Theorem~1.1]{PR19}
and~\cite[Theorem~1.2]{LSS19Ising} give graph-independent open zero-free
neighborhoods of the respective intervals.
For Ising, the reciprocal change of variables converts the bichromatic-edge normalization of~\cite{LSS19Ising} to ours and preserves $I_{\Delta,\delta}$.
Compactness gives a positive tube width for each of these degrees.
Taking the minimum of the corresponding rescaled widths and the large-degree constants proves both lemmas.
\end{proof}

\begin{lemma}\label{lem:potts-truncated-tube}
    For every $a\in(0,1)$ and $\delta>0$ there exists $\rho=\rho(a,\delta)>0$ such that the following holds for all integers $\Delta\ge3$ and $q\ge(2+\delta)\Delta$.
    Let $G=(V,E)$ be a finite simple graph of maximum degree at most $\Delta$.
    Write $m(\sigma)$ for the number of monochromatic edges of $\sigma\in[q]^V$.
    With the convention $0^0=1$, we have
    \[
        Z_G^{\mathrm{Potts}}(w)=\sum_{\sigma\in[q]^V}w^{m(\sigma)}\ne0
        \qquad\text{for all }w\in B([0,1-a],\rho).
    \]
\end{lemma}
The proof of~\Cref{lem:potts-truncated-tube} is deferred to Appendix~\ref{app:potts-tube}.

\begin{proof}[Proof of~\Cref{thm:main}]
For colorings, we may assume $q<n^2/\epsilon$.
Otherwise, at most a fraction $|E|/q\le\epsilon/2$ of all colorings are improper, so returning $q^n$ has relative error at most $\epsilon$.

Define the normalized partition function
\[
    Z(s)=
    \begin{cases}
        Z_G^{\mathrm{hc}}(s\lambda) & \text{for the hardcore model},\\
        2^{-n}Z_G^{\mathrm{Ising}}(1+s(\beta-1)) & \text{for the Ising model},\\
        q^{-n}Z_G^{\mathrm{Potts}}\bigl(\tfrac{99}{100}(1-s)\bigr) & \text{for proper colorings}.
    \end{cases}
\]
By~\Cref{lem:hardcore-tube,lem:ising-tube,lem:potts-truncated-tube}, these polynomials are zero-free on $B([0,1],2\gamma)$ for some $\gamma=\gamma(\delta)>0$.
Indeed, $\Delta\lambda\le12$ and $\Delta|\beta-1|\le6$, so the affine changes of variables pull back the first two tube widths to constants independent of $\Delta$.
Choose the analytic logarithm real at $s=0$ and put $f(s)=n^{-2}\log Z(s)$.
For hardcore and Ising, $Z(0)=1$; for colorings, $(99/100)^{|E|}\le Z(0)\le1$.
Thus $|f(0)|=O(1)$.
Every zero lies at distance at least $\gamma$ from $B([0,1],\gamma)$, so logarithmic differentiation gives
\[
    |f'(s)|\le\frac{\deg Z}{n^2\gamma}\le\frac1\gamma
    \qquad\text{on }B([0,1],\gamma).
\]
Integrating from zero within this convex domain shows that $|f|=O_\delta(1)$ there.

We construct an evaluation oracle for $f$ near $s=0$ using~\Cref{thm:high-temp-spin}.
For the hardcore model, take independent occupied-spin probability $s\lambda/(1+s\lambda)$ and edge weight zero on an occupied pair.
The normalization factor is $(1+s\lambda)^n$, and the left-hand side of~\eqref{eq:high-temp} is at most
\[
    \frac{\Delta s\lambda}{1+s\lambda}\le12s.
\]

For Ising, put $b_s=1+s(\beta-1)$ and $a_s=\max\{1,b_s\}$.
Use uniform fields, divide both edge weights by $a_s$, and restore the factor $a_s^{|E|}$ after evaluation.
The high-temperature bound is
\[
    \frac{\Delta|1-b_s|}{2a_s}\le3s.
\]

For colorings, use uniform fields and monochromatic edge weight $\tfrac{99}{100}(1-s)$.
The high-temperature bound is
\[
    \frac{\Delta(1-\tfrac{99}{100}(1-s))}{q}
    \le\frac{1/100+s}{2}.
\]
All three bounds are at most $1/100$ on the fixed interval $[0,r]$ with $r=1/2400$.
A relative error $u\le1/2$ changes the logarithm by at most $2u$, and all normalization factors are explicit.
We evaluate logarithms by scaling their arguments by powers of two into $[1/2,1]$ and using the series for $\log(1+x)$.
Consequently, \Cref{thm:high-temp-spin} supplies an additive-error oracle for $f$ with cost polynomial in $n,q$ and the inverse requested accuracy.

Apply~\Cref{lem:extrapolation} to accuracy $\epsilon/(2n^2)$.
Computing $\exp(n^2\widehat f(1))$ to relative error $\epsilon/8$ and restoring the factors $2^n$ or $q^n$ gives the desired multiplicative estimate.
The oracle tolerances are $(\epsilon/n)^{O_\delta(1)}$.
Since $q<n^2/\epsilon$ in the coloring case, the total arithmetic cost is $(n/\epsilon)^{O_\delta(1)}$.
\end{proof}

\begin{remark*}[Bit implementation]
The transformed weights have bit length
$O_\delta(b+\log(n/\epsilon))$.
For fixed $\delta$, the rational interpolation nodes and Lagrange
weights have bit length polynomial in $\log(n/\epsilon)$.
Using dyadic oracle outputs at the prescribed accuracy allows exact
rational interpolation.
Together with the bit implementation of~\Cref{thm:high-temp-spin}
and range reduction for logarithms and exponentials, this gives
$(n/\epsilon)^{O_\delta(1)}\operatorname{poly}(b)$ bit operations.
The direct output $q^n$ has the same bound.
\end{remark*}

\section{Counting matchings}\label{sec:matching}

Let $G=(V,E)$ be a simple undirected graph on $n$ vertices. The
monomer--dimer model at activity $\lambda>0$ assigns each matching $M$ of
$G$ probability proportional to $\lambda^{|M|}$, with partition function
\[
 Z_G(\lambda)=\sum_{M\text{ matching in }G}\lambda^{|M|}.
\]
For each $v\in V$, let $p_{G,v}$ be the marginal probability that $v$ is
unmatched. Partitioning the matchings of $G$ according to the edge
covering $v$, if any, gives the monomer
recursion~\cite[Proposition~3.1]{BGKNT07}
\begin{equation}\label{md:delete}
 p_{G,v}=\frac{1}{1+\lambda\sum_{u\in N_G(v)}p_{G-v,u}}.
\end{equation}
For small $\lambda$, expanding~\eqref{md:delete} directly gives
\[
 p_{G,v}=\sum_{k\ge0}\left(-\lambda\sum_{u\in N_G(v)}p_{G-v,u}\right)^k .
\]
The expansion parameter can reach $\lambda\Delta$, where $\Delta$ is the maximum degree of $G$, so convergence in the worst case requires $\lambda<1/\Delta$.
Applying~\Cref{lem:extrapolation} from $r=\Theta(1/\Delta)$ gives running time $n^{2^{O(\Delta)}}$ at fixed $\lambda$ and $\epsilon$, which does not give a quasipolynomial worst-case bound on unrestricted graphs.

To eliminate this $\Delta$ dependence, we use a belief-propagation (BP)
\footnote{Our BP updates one message per vertex using all its neighbors;
standard matching BP updates directed-edge messages with the recipient excluded.}
reference, i.e., the unique fixed point $x \in [0,1]^V$ of the following system, to have a better convergence radius for the recursion.
\begin{equation}\label{md:bp}
    x_v=\frac{1}{1+\lambda\sum_{u\in N_G(v)}x_u}.
\end{equation}
Let $\epsilon_{G,v}=(p_{G,v}-x_v)/x_v$ be the relative difference between $p_{G,v}$ and $x_v$.
Using the same reference vector $x$ for the child errors, the recursion~\eqref{md:delete} gives
\[
 \epsilon_{G,v}
 =\frac{1}{1+\sum_{u\in N_G(v)} \lambda x_u x_v \epsilon_{G-v,u}}-1.
\]
With $\delta_{G,v} \defeq \sum_{u\in N_G(v)}\lambda x_vx_u\,\epsilon_{G-v,u} = x_v/p_{G,v} - 1$
denoting the sum in the denominator, the expansion
$\epsilon_{G,v}=\sum_{j\ge1}(-\delta_{G,v})^j$ converges whenever $\abs{\delta_{G,v}}<1$.
In fact, this holds for all $\lambda<1/2$, a large improvement over $\lambda<1/\Delta$.
To see this, consider the difference between the expected numbers of unmatched vertices in $N_G(v)$ under the monomer--dimer distributions on $G-v$ and $G$.
By linearity of expectation, this difference equals $\sum_{u\in N_G(v)}(p_{G-v,u}-p_{G,u})$.
An alternating-path coupling~\cite{CG24,CFJMYZ25} between the two distributions bounds the absolute difference by $1-p_{G,v}$.
Rewriting~\eqref{md:delete} as $p_{G,v}^{-1}-1=\lambda\sum_{u\in N_G(v)}p_{G-v,u}$ therefore gives
\[
 \left|p_{G,v}^{-1}-1-\lambda\sum_{u\in N_G(v)}p_{G,u}\right|
 \le \lambda(1-p_{G,v}) \le \lambda.
\]
The absolute term $r_v \defeq p_{G,v}^{-1}-1-\lambda\sum_{u\in N_G(v)}p_{G,u}$ satisfies $\abs{r_v}\le\lambda$. Rearranging gives
\begin{align*}
    \frac{x_v}{p_{G,v}} = x_v (1+r_v) + \sum_{u \in N_G(v)} \lambda x_u x_v \frac{p_{G,u}}{x_u}.
\end{align*}
Note that $x_v+\sum_{u\in N_G(v)}\lambda x_ux_v=1$.
Thus, $x_v/p_{G,v}$ is a convex combination of $1+r_v$ and $p_{G,u}/x_u$ for $u\in N_G(v)$, with positive weight $x_v$ on $1+r_v$.
For $0<\lambda<1$, if the largest of these ratios and their reciprocals exceeded $1/(1-\lambda)$, this identity would force a still larger one, since $1+r_v\in[1-\lambda,1+\lambda]$.
Consequently, $x_v/p_{G,v}\in[1-\lambda,1/(1-\lambda)]$, which establishes the desired $\abs{\delta_{G,v}}<1$ when $\lambda<1/2$.

Applying the framework of~\Cref{sec:high-temp} to this BP-centered expansion yields a fast algorithm for approximating the monomer marginals and hence $\log Z_G(\lambda)$.
To control deletion effects and terminal errors at fixed accuracy, we use a weighted stopping rule under the stronger condition $\lambda \lesssim 1/\log n$, giving the low-activity oracle stated in~\Cref{md:oracle} below.
Combining this oracle with the stability of the matching generating polynomial~\cite{HL72} then yields the quasipolynomial-time approximation algorithm in~\Cref{thm:match}.

\begin{proposition}\label{md:oracle}
Let $G=(V,E)$ be a finite simple graph with $n\ge2$ vertices, let $\eta\in(0,1/2)$, and set $\ell=\lceil\log_2(2n/\eta)\rceil$.
For every activity $\lambda\in(0,1]$ satisfying the high-temperature condition $\lambda\ell\le 1/2048$, there is a deterministic algorithm approximating $\log Z_G(\lambda)$ to additive error $\eta$ using $\exp(O(\ell\log n))$ arithmetic operations.
\end{proposition}

\subsection{Algorithm and analysis}\label{subsec:matching-alg}
By self-reducibility, it suffices to estimate the monomer marginals along a
vertex-deletion order to relative error $\eta/(2n)$.
After approximating the BP fixed point, we bound compression, stopping,
and truncation errors separately.

\begin{lemma}\label{lem:md-reference}
For $0<\lambda\le1$, we can compute an estimate $x$ of the fixed point $x^*$
in~\eqref{md:bp} using $\operatorname{poly}(n)$ arithmetic operations and
comparisons. The estimate satisfies the following for every $v\in V$.
\begin{enumerate}
 \item $1/n\le x_v\le1$.
 \item $0\le\log(x_v^*/x_v)\le1/(128n)$.
 \item $0\le c_v:=x_v^{-1}-1-\lambda\sum_{u\in N_G(v)}x_u\le1/64$.
\end{enumerate}
\end{lemma}
\begin{proof}
The BP update $T$ preserves $[1/n,1]^V$, reverses coordinatewise order,
and contracts the sup norm in log coordinates by a factor of at most
$1-1/n$, since the absolute Jacobian row sums are $1-T_v(x)\le1-1/n$.
Starting from the box endpoints, iterate $(l,u)\gets(T(u),T(l))$ with
outward rounding on a grid of step $1/(1024n^3)$, which preserves
$l\le x^*\le u$ and gives $b_{\rm new}\le(1-1/n)b+1/(512n^2)$ for
$b=\max_v\log(u_v/l_v)$.
After polynomially many rounds, $b\le1/(128n)$, and taking $x=l$
gives the desired bounds: the enclosure $e^{-b}x^*\le x\le x^*$
and the fixed-point equation imply
$0\le c_v\le (1-e^{-2b})x_v^{-1}\le2nb\le1/64$.
Evaluating $T$ and rounding on the polynomial-size grid take polynomially
many operations and comparisons per round.
\end{proof}

Fix a root $v_0$ in the current graph $G$, with $n$ still denoting
the original graph size, and take $x$ from~\Cref{lem:md-reference}.
A valid deletion history is a walk from $v_0$ without repeated vertices.
As in~\Cref{subsec:alg}, the dynamic programming table has $n+1$ rows indexed by
history depth $d\in\{0,\ldots,n\}$ and columns indexed by the final augmented
state $a=(v,t)\in V\times\{0,\ldots,128n\ell\}$. For a history
$h=(v_0,\ldots,v_d)$, write $v=v_d$ and $S_j=(v_j,t_j)$ for $0\le j\le d$,
where $S_d=(v,t)$ and the counter $t_j$ tracks the capped prefix weight via
\[
 t_0=0,\qquad t_{j+1}=\min\{128n\ell,\,t_j+w_{v_j}\},
 \qquad w_v=\lceil nx_v\rceil.
\]
Each entry stores an approximation to the relative correction $g_h(s)$
defined below, represented by a weighted sum of products
\[
 \widehat g_h(s)=\sum_i a_i(s)\prod_{j=1}^d f_{i,j}(S_{j-1},S_j),
 \qquad |f_{i,j}|\le1,\qquad \sum_i\|a_i\|_\rho\le A:=1/(64n),
\]
where each $a_i(s)$ has degree at most $m$, $\rho=1/(512n^2)$, and
$\|P\|_\rho=\sum_j\rho^j|[s^j]P|$.

For a valid history $h=(v_0,\ldots,v_d)$ with final state $S_d=(v,t)$,
we initialize the corresponding monomer message $p_h(s)$ to the
constant $x_v$ when $t=128n\ell$.
On every other valid history, we let
\begin{equation}\label{md:homotopy}
 p_h(s)=\left(1+(1-s)C_h+
       \lambda\sum_{\substack{u\sim v\\u\ne v_j\ \forall j<d}}
       p_{h\parallel u}(s)\right)^{-1},
\end{equation}
where $C_h=c_v+\lambda\sum_{j<d}x_{v_j}\mathbf1_{\{v_j\sim v\}}$ uses
the residual $c_v=x_v^{-1}-1-\lambda\sum_{u\in N_G(v)}x_u$ from
\Cref{lem:md-reference} and $h\parallel u$ extends $h$ by $u$.
The recursion~\eqref{md:homotopy} interpolates between $p_h(0)=x_v$ and $q_h:=p_h(1)$,
the monomer recursion with boundary values $x_v$ at early-stopped histories.
Rewriting~\eqref{md:homotopy} as a recurrence for the relative correction
$g_h(s)=p_h(s)/x_v-1$ and truncating at degree $m$
gives the idealized recursion for the dynamic programming table:
\begin{equation}\label{md:series}
 g_h(s)=\frac{U_h(s)}{1-U_h(s)}
       \equiv\sum_{j=1}^m U_h(s)^j\pmod{s^{m+1}},
\end{equation}
where $U_h(s)=s x_vC_h-\sum_{u\sim v}\lambda x_vx_u I_u(h)g_{h\parallel u}(s)$
and $I_u(h)=\prod_{j<d}\mathbf1_{\{v_j\ne u\}}$.

Given $m\ge1$ and error parameter $\xi\in(0,1]$, we implement~\eqref{md:series} via~\Cref{alg:compress-suffix}. Initialize row $n$ and all
stopped cells with $t=128n\ell$ to zero and then fill rows $d=n-1,\ldots,0$ backwards: at an
unstopped cell $(d,a)$ with $a=(v,t)$, set $t'=\min\{128n\ell,\,t+w_v\}$
and form the list
\[
\begin{aligned}
 \widehat U_{d,a}(S_0,\ldots,S_d;s)
 &= s x_v\left(c_v+\lambda\sum_{j=0}^{d-1}x_{v_j}\mathbf1_{\{v_j\sim v\}}\right) \\
 &\quad - \sum_{u\sim v}\lambda x_vx_u\left(\prod_{j=0}^{d-1}\mathbf1_{\{v_j\ne u\}}\right)
   \widehat g_{d+1,(u,t')}(S_0,\ldots,S_d,(u,t');s),
\end{aligned}
\]
absorbing each child product's last transition factor into its coefficient
and combining local factors positionwise. We then compute
\[
 T_0=0,\qquad
 T_j=\mathcal C\bigl([\widehat U_{d,a}(1+T_{j-1})]_{\le m}\bigr)\quad(1\le j\le m),\qquad
 \widehat g_{d,a}=T_m,
\]
where $\mathcal C$ compresses degreewise to $M=\lceil 4n^2/\xi^2\rceil$
terms using~\Cref{lem:compress} under $\mu_{d,a}$. Finally, return
$\widehat p=x_{v_0}\bigl(1+\widehat g_{0,(v_0,0)}((v_0,0);s)\bigr)$.

Here $\mu_{d,a}=\operatorname{Law}((S_0,\ldots,S_d)\mid S_d=a)$ is
defined by the Markov chain $(S_j)_{j\ge0}$ starting at $S_0=(v_0,0)$ that
moves from $(v,t)$ to $(u,t')$ for $u\sim v$ and
$t'=\min\{128n\ell,\,t+w_v\}$ with probability $P_{vu}/P_v$ when $P_v>0$,
and stays at $(v,t)$ when $P_v=0$, where $P_{vu}=\lambda x_vx_u$ and
$P_v=\sum_{u\sim v}P_{vu}=1-x_v-x_v c_v$. We compute only cells with
$\Pr(S_d=a)>0$; every child with $P_{vu}>0$ also has positive probability.

\begin{lemma}\label{md:local}
For a given vertex $v$, an integer $m\ge1$, and $0<\xi\le1$, the algorithm
above returns a polynomial $\widehat p$ of degree at
most $m$ satisfying $\|\widehat p-[p_{(v)}]_{\le m}\|_\rho\le\xi$ using
$\operatorname{poly}(n,\ell,m,\xi^{-1})$ arithmetic operations and
comparisons, where $[P]_{\le m}$ denotes truncation after degree $m$.
\end{lemma}
\begin{proof}
Take $v_0=v$. Since histories of the Markov chain $(S_j)_{j\ge0}$ may
revisit vertices, we extend $g_h$ to all histories of the chain via the
backward recursion~\eqref{md:series} with the same initialization. On valid histories, the indicator
$I_u(h)$ zeroes out previously visited neighbors just as in~\eqref{md:homotopy},
so this extension agrees with $g_h(s)=p_h(s)/x_v-1$, in particular at
the root.

We next show by backward induction that every stored list
has weight $\sum_i\|a_i\|_\rho\le A=1/(64n)$. This holds at row $n$ and at
stopped cells. If child lists have weight at most $A$, then since
$c_v+\lambda d\le n$, $\rho n=A/8$, and $P_v\le1-x_v$, the list for
$\widehat U_{d,a}$ has weight at most
\[
 \rho x_v(c_v+\lambda d)+AP_v\le \frac{x_v A}{8}+A(1-x_v)=A(1-7x_v/8).
\]
Since truncated multiplication modulo $s^{m+1}$ is submultiplicative and
compression does not increase list weight, each iterate $T_j$ has weight
at most $A(1-7x_v/8)(1+A)\le A$, using $A=1/(64n)<7/(8n)\le7x_v/8$.
This completes the induction. Since $|f_{i,j}|\le1$, we get
$\|\widehat g_{d,a}(h)\|_\rho\le A$, and the same induction in the
pointwise norm gives $\|[g_h]_{\le m}\|_\rho\le A$ as well as
$\|\widehat U_{d,a}\|_\rho,\|U_h\|_\rho\le A$.

It remains to analyze the time complexity and error of this algorithm.
There are $O(n^3\ell)$ cells, and each performs $m$ compressions on
polynomial-size lists with at most $mM$ terms per output list. The
pointwise product of two terms factors along the history, so conditional
inner products under $\mu_{d,a}$ are computed by a forward pass along the
Markov chain $(S_j)_{j\ge0}$, normalized by $\Pr(S_d=a)$. Since
$M=O(n^2/\xi^2)$, the total cost is
$\operatorname{poly}(n,\ell,m,\xi^{-1})$ arithmetic operations and
comparisons.

For cells of zero probability, set $E_{d,a}=0$; otherwise let
\[
 E_{d,a} \defeq \left(\mathbb E_{h\sim\mu_{d,a}}
   \bigl[\|e_{d,a}(h;s)\|_\rho^2\bigr]\right)^{1/2},
\]
where $e_{d,a}(h;s)=\widehat g_{d,a}(h;s)-[g_h(s)]_{\le m}$. Writing
$\mu_d=\operatorname{Law}(S_0,\ldots,S_d)$ for the marginal distribution
of the first $d$ steps of the chain, we claim that the row error
\[
 E_d \defeq \left(\sum_a\Pr(S_d=a)E_{d,a}^2\right)^{1/2}
     = \left(\mathbb E_{h\sim\mu_d}\bigl[\|e_{d,S_d}(h;s)\|_\rho^2\bigr]\right)^{1/2}
\]
satisfies
\[
 E_n=0,\qquad E_d\le \frac{\xi}{2n}+E_{d+1}\quad (0\le d<n).
\]

The initialization gives $E_n=0$. Set $\widetilde g_{d,a}=0$ at stopped
cells; at an unstopped cell $(d,a)$ with $d<n$, define
the uncompressed iterates on the stored children by $T_0^\circ=0$ and
$T_j^\circ=[\widehat U_{d,a}(1+T_{j-1}^\circ)]_{\le m}$, and write
$\widetilde g_{d,a}(h;s)=T_m^\circ=[\widehat U_{d,a}/(1-\widehat U_{d,a})]_{\le m}$
for the exact update~\eqref{md:series} modulo $s^{m+1}$ on the stored
children. By the triangle inequality,
\[
 E_d \le \left(\mathbb E_{h\sim\mu_d}
   \bigl[\|\widehat g_{d,S_d}(h)-\widetilde g_{d,S_d}(h)\|_\rho^2\bigr]\right)^{1/2}
   +\left(\mathbb E_{h\sim\mu_d}
   \bigl[\|\widetilde g_{d,S_d}(h)-[g_h]_{\le m}\|_\rho^2\bigr]\right)^{1/2},
\]
splitting into the local error and the error propagated from the children.

To bound the local error, consider one compression $\widehat F=\mathcal C F$
at cell $(d,a)$ whose degree-$j$ input list has weight $B_j$, with total
weight $B=\sum_{j=1}^m\rho^j B_j\le A$. Applying~\Cref{lem:compress} at
each degree and Minkowski's inequality gives
\[
 \left(\mathbb E_{h\sim\mu_{d,a}}
   \bigl[\|\widehat F(h;s)-F(h;s)\|_\rho^2\bigr]\right)^{1/2}
 \le \sum_{j=1}^m\rho^j\bigl\|[s^j](\widehat F-F)\bigr\|_{L^2(\mu_{d,a})}
 \le \frac{B}{\sqrt M}\le \frac{A\xi}{2n}\le \frac{\xi}{4n}.
\]
Let $\delta_j$ be the $L^2(\mu_{d,a})$ error of $T_j$ relative to
$T_j^\circ$. Since $\|\widehat U_{d,a}\|_\rho\le A$, multiplication by
$\widehat U_{d,a}$ scales the previous error by at most $A$ and
compression adds at most $\xi/(4n)$, giving
\[
 \delta_0=0,\qquad \delta_j\le A\delta_{j-1}+\frac{\xi}{4n},
\]
whence the local error in cell $(d,a)$ is
$\delta_m\le\frac{\xi}{4n(1-A)}\le\frac{\xi}{2n}$.

For the propagated error, subtract the two rational updates
in~\eqref{md:series} modulo $s^{m+1}$. Since
$\|U_h\|_\rho,\|\widehat U_{d,a}\|_\rho\le A$, each inverse factor has
$\rho$-norm at most $(1-A)^{-1}$, so for any history $h$ in the
support of $\mu_{d,a}$ with $a=(v,t)$ and $t'=\min\{128n\ell,\,t+w_v\}$,
\[
\begin{aligned}
 \bigl\|\widetilde g_{d,a}(h)-[g_h]_{\le m}\bigr\|_\rho
 &\le \frac{1}{(1-A)^2}\sum_{u\sim v}P_{vu}
    \bigl\|e_{d+1,(u,t')}(h\parallel u;s)\bigr\|_\rho \\
 &\le \mathbb E\Bigl[
    \bigl\|e_{d+1,S_{d+1}}(h\parallel S_{d+1};s)\bigr\|_\rho
    \,\Bigm|\,(S_0,\ldots,S_d)=h\Bigr],
\end{aligned}
\]
where the second step uses $P_v\le1-x_v\le(1-A)^2$, which follows from
$x_v\ge1/n>2A$ by~\Cref{lem:md-reference}. By Jensen's inequality and
averaging over $\mu_d$, the propagated error is at most $E_{d+1}$,
proving the recurrence. Iterating gives $E_0\le\xi/2$; since the root
history is deterministic and $x_{v_0}\le1$, we conclude
$\|\widehat p-[p_{(v)}]_{\le m}\|_\rho=x_{v_0}E_0\le\xi/2<\xi$.
\end{proof}

\begin{lemma}\label{md:cutoff}
For $h=(v_0,\ldots,v_d)$, let $p_h=p_{G-\{v_j:j<d\},v_d}$ be the exact
monomer message. The root satisfies
$|\log p_{(v_0)}-\log q_{(v_0)}|<\eta/(8n)$.
\end{lemma}
\begin{proof}
At an unstopped history, $\sum_{j<d}x_{v_j}\le t_d/n<128\ell$.
Hence $C_h\le1/64+128\lambda\ell\le1/8$.
At stopped leaves $q_h=x_v$. Comparing the denominators of $q_h$ and
$x_v$ by induction gives $x_v/2\le q_h\le2x_v$, since the constants
$1$ and $1+C_h$ are within a factor of two. Write
$E_h=|p_h-q_h|/\sqrt{p_hq_h}$. At an internal node, subtraction of
the two monomer recursions and Cauchy--Schwarz give
\[
 E_h\le\sqrt{(1-p_h)(1-q_h)}\,\max_u E_{h\parallel u}
 \le e^{-q_h/2}\max_u E_{h\parallel u}
 \le e^{-w_v/(8n)}\max_u E_{h\parallel u}.
\]
The last inequality uses $w_v\le 2nx_v\le4nq_h$.
Since $t_{d+1}\le t_d+w_v$, the quantity $e^{-t_d/(8n)}E_h$ is at most
its maximum over the children. At an unstopped leaf $E_h=0$, and at a
stopped leaf $E_h\le2n$ since both messages lie in $[1/n,1]$.
Thus the root error is at most $2ne^{-16\ell}<\eta/(8n)$, using
$|\log p_h-\log q_h|\le E_h$.
\end{proof}

\begin{proof}[Proof of \Cref{md:oracle}]
Use the original $n$ and $\eta$ throughout the deletion order.
For $|s|\le2$, the constant term $1+(1-s)C_h$ in \eqref{md:homotopy}
has real part at least $7/8$. Since inversion preserves the open right
half-plane, induction gives holomorphy on $|s|<2$ and $|p_h(s)|\le8/7$.

At each root, set $m=16\ell$ and use \Cref{md:local} with tolerance
$\xi=\eta\rho^m/(128n^2)$, rounded down to a power of two. Evaluate
the resulting polynomial $\widehat p$ at $s=1$. The coefficient error
is amplified by at most $\rho^{-m}$, while Cauchy's estimate on the
circle of radius $3/2$ bounds the tail. Thus
\[
 \bigl|\widehat p(1)-q_{(v_0)}\bigr|
 \le \rho^{-m}\xi+\frac{24}{7}(2/3)^{m+1}
 \le \frac{\eta}{64n^2}.
\]
Since $q_{(v_0)}\ge1/(2n)$, the error in the logarithm is at most
$\eta/(8n)$. The cutoff and numerical logarithm evaluation each add
at most $\eta/(8n)$, so summing over the deletion order gives the
claimed accuracy. There are $n$ root computations with
$\log(1/\xi)=O(\ell\log n)$, giving the stated arithmetic bound.
\end{proof}

\subsection{Analytic extrapolation}
As in~\Cref{sec:extrapolation}, we bound the interpolation error with
exact samples and the amplification of the sampling error. Here the
activity extrapolation uses a single interpolation on a disk.

\begin{lemma}\label{md:lagrange}
Let $f$ be holomorphic on a neighborhood of $|w|\le r$ and
satisfy $|f|\le M$ on its boundary. Let $0<a<x<r$ and $k\ge1$.
The polynomial interpolating values of $f$ at $ja/k$,
$0\le j\le k$, each with additive error at most $\eta$, satisfies
\[
 |f(x)-\widehat P(x)|\le
 \frac{rM}{r-x}\left(\frac{x}{r}\right)^{k+1}
 +\left(\frac{2\mathrm e x}{a}\right)^k\eta .
\]
\end{lemma}
\begin{proof}
Write $u_j=ja/k$. For the exact-data error integrate
$\frac{f(w)}{w-x}\prod_{j=0}^k\frac{x-u_j}{w-u_j}$
around the circle and divide by $2\pi i$. Each quotient has
modulus at most $(x-u_j)/(r-u_j)\le x/r$.
The sum of the absolute Lagrange weights is at most
$(kx/a)^k\sum_{j=0}^k [j!(k-j)!]^{-1}\le (2\mathrm e x/a)^k$.
\end{proof}

\begin{proof}[Proof of \Cref{thm:match}]
At $\lambda=0$ we return $Z_G(0)=1$ in $O(1)$ arithmetic operations. For $\lambda>n$ we compute
the partition function exactly by evaluating
\[
 Z_{G[S]}(\lambda)=Z_{G[S\setminus\{v\}]}(\lambda)
   +\lambda\sum_{u\in N_G(v)\cap S}Z_{G[S\setminus\{u,v\}]}(\lambda)
\]
for every nonempty $S\subseteq V(G)$, choosing any $v\in S$.
With $Z_{G[\varnothing]}(\lambda)=1$, this takes $O(n2^n)$
arithmetic operations, within the claimed bound.

Now suppose $0<\lambda\le n$. By decreasing $\epsilon$ to the largest
power of two at most $\min(\epsilon,1/2)$, assume $\epsilon$ has this form.
The zeros of $Z_G(z)$ are strictly negative reals by the
Heilmann--Lieb theorem~\cite{HL72}.
Let $B=\lceil\lambda\rceil$, $L=\lceil\log_2(2n/\epsilon)\rceil$,
and $t=2^{-20}/(BL)$. Set
\[
 z_0=\lambda t^2,\qquad
 z(w)=z_0\left(\frac{1+w}{1-w}\right)^2,\qquad
 b=\frac{1-t}{1+t},\qquad r=\frac1{1+t}.
\]
Then $z(b)=\lambda$, and $z(w)\le2z_0$ on the sampling interval $[0,1/32]$.
For each zero $-\rho_i$ of $Z_G$, the normalized factor becomes
\[
 \frac{z(w)+\rho_i}{z_0+\rho_i}
 =\frac{(1-\omega_i w)(1-\overline{\omega_i} w)}{(1-w)^2},
 \qquad |\omega_i|=1,
\]
since its numerator roots satisfy $(1+w)/(1-w)=\pm i\sqrt{\rho_i/z_0}$.
Thus $F(w)=\log Z_G(z(w))$, real at the origin, is holomorphic on the
unit disk. Since $z_0\le1$, we have $Z_G(z_0)\le(n+1)^n$. Taylor
expansion of the logarithms, using at most $n/2$ zeros and $B\le n$, gives
\[
 \max_{|w|=r}|F(w)|\le n\log(n+1)+2n\log((1+t)/t)\le20nL.
\]

Interpolate $F$ from $[0,1/32]$ to $b$ with degree $K=16L/t$ and
sample tolerance $\eta=(\epsilon/8)256^{-K}$. Since $b/r=1-t$ and
$r/(r-b)=1/t$, the two terms in \Cref{md:lagrange} are bounded by
\[
 20nL t^{-1}e^{-tK}\le\epsilon/8,
 \qquad (64\mathrm e)^K\eta\le\epsilon/8.
\]
For this tolerance, $\ell=L+3+8K\le16K$. At every node
$w_j=j/(32K)$ we have $z(w_j)\ell\le32Bt^2K=1/2048$, so
\Cref{md:oracle} supplies the samples. Exponentiating the estimate
of $F(b)$ and rounding to relative error $\epsilon/8$ gives the claimed
multiplicative accuracy.

There are $K+1=O(BL^2)$ samples of precision $\ell=O(BL^2)$.
Applying \Cref{md:oracle} at this precision gives the arithmetic cost
$\exp(O(BL^2\log(2n)))$.
%
\end{proof}

\begin{remark*}[Bit implementation]
Assume rational numerical inputs of at most $b$ bits each.
For $0<\lambda\le n$, after each compression in~\Cref{md:local},
round the local factors and weighted coefficients $\rho^j[s^j]a_i$
toward zero on dyadic grids of mesh at most $1/(4nM)$ and $1/(4mM^2)$,
respectively. This preserves the weight bound and adds at most $1/(4M)$
for each rounding, giving total compression and rounding error at most $A/\sqrt M+1/(2M)\le\xi/(4n)$.
Since $\log(1/\xi)=O(K\log n)$, the common-denominator argument
from~\Cref{sec:high-temp} gives polynomial bit length in $n,K,b$
for the stored numbers, conditional inner products and interpolation data.
Together with range reduction for logarithms and exponentials,
this gives
$\exp(O(BL^2\log(2n)))\operatorname{poly}(b)$ bit operations.

For $\lambda=p/q>n$ with positive integers $p,q$, the scaled subset
entries $q^{|S|}Z_{G[S]}(p/q)$ are integers bounded by
$(n+1)^n\max\{p,q\}^n$.
Thus this branch takes $2^n\operatorname{poly}(n,b)$ bit operations,
within the claimed bound.
\end{remark*}

\bibliographystyle{alpha}
\bibliography{refs}

\clearpage
\appendix
\section{Zero-freeness for the antiferromagnetic Potts model}\label{app:potts-tube}

\begin{proof}[Proof of~\Cref{lem:potts-truncated-tube}]
Fix $\delta>0$, $a\in(0,1)$, and a real $w\in[0,1-a]$. We prove nonvanishing on a disk about $w$ whose radius depends only on $\delta$ and $a$. We use the single-vertex log-ratio recursion for the Potts model with pinned leaves~\cite{BBR26}.

\emph{Boundary conditions.}
An instance $H$ consists of a graph on free vertices and, at each vertex $v$, counts $c_{v,k}$ of incident leaves pinned to color $k$. Write $d_H(v)$ for its number of free neighbors and require $d_H(v)+\sum_k c_{v,k}\le\Delta$. Its partition function is
\begin{align*}
    Z_H(z)=\sum_{\sigma:V(H)\to[q]}
       z^{m_H(\sigma)+\sum_{v\in V(H)} c_{v,\sigma(v)}},
\end{align*}
where $m_H$ counts monochromatic edges between free vertices. Let $Y_{v,k}(z)$ be the restricted partition function with $v$ fixed to $k$ after removing the pins incident to $v$. Equivalently, delete $v$ and attach a pin of color $k$ to each former free neighbor.
Each such neighbor loses one free edge and gains one pin, so the degree constraint is preserved.
Thus $Y_{v,k}$ is the partition function of a smaller admissible instance, and
\begin{align}\label{eq:potts-reinsert-pins}
    Z_H(z)=\sum_{k=1}^q z^{c_{v,k}}Y_{v,k}(z).
\end{align}
At $w$ all these partition functions are positive. Indeed, greedily coloring the free vertices while avoiding their pinned colors always leaves at least $q-\Delta$ choices. With the pins at $v$ removed, its real marginal is $\nu_k=Y_{v,k}(w)/\sum_\ell Y_{v,\ell}(w)$. Conditional on all other spins, at least $q-d_H(v)$ colors at $v$ have weight one and each weight is at most one. Averaging gives
\begin{align}\label{eq:potts-depinned-bound}
    \nu_k\le (q-d_H(v))^{-1}\le 2/q .
\end{align}

\emph{One step of the recursion.}
Consider a free vertex with $d$ free neighbors, pin counts $\mathbf c=(c_1,\ldots,c_q)$, and marginal $\nu$ after removing those pins. For $b=0$ or $b\in[q]$ define
\begin{align*}
    S_b(z,\mathbf u)=\sum_{k=1}^q \nu_k e^{u_k} z^{c_k+\mathbf 1_{\{k=b\}}}.
\end{align*}
The index $0$ denotes the original collection of pins. For $b\in[q]$ we add one pin and require $d+\sum_k c_k+1\le\Delta$. For distinct $i,j\in[q]$ put
\[
    F(z,\mathbf u)=\log S_i(z,\mathbf u)-\log S_j(z,\mathbf u),
    \qquad \operatorname{osc}(\mathbf u)=\max_{k,\ell}|u_k-u_\ell|.
\]
We will use the following bound for some $r_a,C_a>0$ depending only on $a$. If $|z-w|,\|\mathbf u\|_\infty\le r_a$, then
\begin{align}\label{eq:potts-local-perturbation}
    |F(z,\mathbf u)-F(w,\mathbf 0)|
    \le \frac{\operatorname{osc}(\mathbf u)}{q-\Delta}
       +\frac{C_a}{q}\bigl(|z-w|+\|\mathbf u\|_\infty^2\bigr).
\end{align}

To prove the linear term, let $p^{(b)}$ be the real marginal after the pin of color $b$ is added. Differentiation gives
$D_{\mathbf u}F(w,\mathbf 0)[\mathbf u]=\sum_k(p^{(i)}_k-p^{(j)}_k)u_k$.
Suppose first that $S_i(w,\mathbf 0)\le S_j(w,\mathbf 0)$. At every coordinate $k\ne i$, the numerator defining $p^{(i)}_k$ is at least the one defining $p^{(j)}_k$, so $p^{(i)}_k\ge p^{(j)}_k$. The total variation distance is consequently at most $p^{(j)}_i\le1/(q-\Delta)$, by conditioning on the other spins with all pins present. Interchanging $i,j$ handles the other ordering. Pairing the positive and negative masses of $p^{(i)}-p^{(j)}$ now gives
\[
    |D_{\mathbf u}F(w,\mathbf 0)[\mathbf u]|
       \le \frac{\operatorname{osc}(\mathbf u)}{q-\Delta}
\]
also for complex $\mathbf u$.

For the remainder, at most $\Delta-d$ colors have a positive exponent in $S_b$, for each allowed $b$ including $0$. By~\eqref{eq:potts-depinned-bound},
\[
    S_b(w,\mathbf 0)\ge 1-\frac{\Delta-d}{q-d}\ge\frac12 .
\]
Choose $0<R_a<a/2$ sufficiently small. On $|z-w|,\|\mathbf u\|_\infty\le R_a$ we have $|z|\le1-a/2$ and, for every $m\ge1$,
\[
    |z^m-w^m|\le m(1-a/2)^{m-1}|z-w|\le (2/a)^2 |z-w|.
\]
Since $\sum_k\nu_k=1$, reducing $R_a$ to ensure $e^{R_a}R_a+(2/a)^2R_a<1/4$ gives $\Re S_b(z,\mathbf u)>1/4$ throughout this domain. Use the principal logarithms there. The difference
\[
    S_i-S_j=(z-1)\bigl(\nu_i e^{u_i}z^{c_i}-\nu_j e^{u_j}z^{c_j}\bigr)
\]
has modulus $O(1/q)$ by~\eqref{eq:potts-depinned-bound}. Integrating $1/\zeta$ along the segment between the two sums gives $|F|=O(1/q)$. Take $r_a=R_a/4$. Cauchy's estimate in $z$ bounds $|F(z,\mathbf u)-F(w,\mathbf u)|$ by $C_a|z-w|/q$ on the smaller domain. For $\mathbf u\ne\mathbf 0$, apply the Taylor coefficient bounds on $|\lambda|=R_a$ to $F(w,\lambda\mathbf u/\|\mathbf u\|_\infty)$. The sum of terms of degree at least two at $\lambda=\|\mathbf u\|_\infty$ is at most $C_a\|\mathbf u\|_\infty^2/q$. These estimates prove~\eqref{eq:potts-local-perturbation}.

\emph{Induction on the number of free vertices.}
Put $g=\delta/(1+\delta)$. Choose $0<E\le r_a$ and $0<h\le r_a$ so that $C_aE^2\le gE/3$ and $C_ah\le gE/3$. We show that $Z_H$ is nonzero on $B(w,h)$ and that for every $v$ the analytic logarithms
\[
    u_{v,k}(z)=\log\left(\frac{Y_{v,k}(z)/Y_{v,1}(z)}
                                  {Y_{v,k}(w)/Y_{v,1}(w)}\right),
    \qquad u_{v,k}(w)=0 ,
\]
satisfy $\operatorname{osc}(\mathbf u_v(z))\le E$ there. The empty instance has partition function one. At the induction step all $Y_{v,k}$ are nonzero on the disk by the smaller-instance assertion, so these logarithms exist. In particular, $u_{v,1}=0$.

Fix distinct colors $i,j$ and order the free neighbors of $v$ as $x_1,\ldots,x_d$. In the representation of $Y_{v,j}$ by deletion, change the added pins from color $j$ to $i$ in this order. For step $\ell$, let $H_\ell$ be the instance on $H-v$ with the new pins of color $i$ at the earlier neighbors, pins of color $j$ at the later ones, and no new pin at $x_\ell$. This leaves room for one pin at $x_\ell$. Use its pin counts and its real marginal with the pins at $x_\ell$ removed to form the local function $F_\ell$. Let $\mathbf u_\ell$ denote the log-ratio vector at $x_\ell$ in $H_\ell$. By~\eqref{eq:potts-reinsert-pins}, the ratio for this pin change at $z$ is $S_i(z,\mathbf u_\ell(z))/S_j(z,\mathbf u_\ell(z))$, since the common factor from the reference color cancels. Telescoping gives
\[
    u_{v,i}(z)-u_{v,j}(z)
       =\sum_{\ell=1}^d\bigl(F_\ell(z,\mathbf u_\ell(z))-F_\ell(w,\mathbf 0)\bigr).
\]
The logarithms on both sides agree at $w$ and their exponentials agree on the disk. By induction $\|\mathbf u_\ell\|_\infty\le\operatorname{osc}(\mathbf u_\ell)\le E$. Hence~\eqref{eq:potts-local-perturbation} gives
\[
    |u_{v,i}(z)-u_{v,j}(z)|
       \le \frac{dE}{q-\Delta}+\frac{C_a d}{q}(h+E^2)
       \le (1-g)E+C_a(h+E^2)<E .
\]
This also covers $d=0$, when the sum is empty. Finally, for the original pins at $v$, equation~\eqref{eq:potts-reinsert-pins} becomes
\[
    Z_H(z)=\frac{Y_{v,1}(z)}{Y_{v,1}(w)}
            \left(\sum_k Y_{v,k}(w)\right) S_0(z,\mathbf u_v(z)).
\]
The first two factors are nonzero and $\Re S_0>1/4$, which proves the induction claim. Applying it with no pins and taking the union of $B(w,h)$ over $w\in[0,1-a]$ proves the lemma with $\rho=h$.
\end{proof}

\end{document}